\def\mycmd{2}

\if\mycmd1
\documentclass[11pt,onecolumn, draftcls]{IEEEtran}
\else 
\documentclass[lettersize,journal]{IEEEtran}
\fi
\usepackage[latin9]{inputenc}
\usepackage{tabularx}
\usepackage{fontawesome}
\usepackage{threeparttable}

\usepackage{colortbl}    % 표 색상 (선 색상 변경)
\usepackage{pifont}     % 기호 (\ding{})

\usepackage[british]{babel}
\usepackage{multicol}
\usepackage{array,ragged2e}
\usepackage{float}
\usepackage{mathtools}
\usepackage{amsmath}
\usepackage{amsthm}
\usepackage{amssymb}
\usepackage{graphicx}
\usepackage{wasysym}
\usepackage{setspace}
\usepackage{color}
\usepackage{bm}
\usepackage{cases}
\usepackage{makecell}
\usepackage{tikz}
\usepackage{flowchart}
\usepackage{amsfonts}
\usepackage{steinmetz}
\usetikzlibrary{matrix,shapes,arrows,positioning,chains}
\usepackage{lmodern,babel,adjustbox,booktabs,multirow}
\usepackage{makecell}
\usepackage{pbox}
\usepackage{epsfig}
\usepackage{subfigure}
\usepackage{tcolorbox}
\usepackage{enumitem}

\allowdisplaybreaks[4]

\makeatletter
\newcommand{\multiline}[1]{%
  \begin{tabularx}{\dimexpr\linewidth-\ALG@thistlm}[t]{@{}X@{}}
    #1
  \end{tabularx}
}

\floatstyle{ruled}
\newfloat{algorithm}{tbp}{loa}
\providecommand{\algorithmname}{Algorithm}
\floatname{algorithm}{\protect\algorithmname}

\theoremstyle{plain}

\theoremstyle{plain}

\theoremstyle{plain}

\usepackage{epsfig}
\usepackage[caption=false,font=normalsize,labelfont=sf,textfont=sf]{subfig}
\usepackage{cite}
\usepackage{stfloats}
\usepackage{graphicx}
\usepackage{multirow}
\usepackage{array}
\usepackage{enumerate}
\usepackage{graphicx}
\usepackage{times}
\usepackage{algorithm}
\usepackage{algpseudocode}
\theoremstyle{remark}

\DeclareMathOperator*{\argmin}{arg\,min}
\DeclareMathOperator*{\argmax}{arg\,max}
\makeatother

\algrenewcommand\algorithmicindent{1.0em}%
\providecommand{\lemmaname}{Lemma}
\providecommand{\propositionname}{Proposition}

\providecommand{\theoremname}{Theorem}
\providecommand{\theoremname}{Definition}
\newcommand{\rom}[1]{\uppercase\expandafter{\romannumeral #1\relax}}

\DeclarePairedDelimiter\ceil{\lceil}{\rceil}

\newcounter{problem}
\newcounter{save@equation}
\newcounter{save@problem}

\definecolor{lightergray}{gray}{0.9}
\definecolor{ForestGreen}{RGB}{34,139,34}  % 녹색 정의

\newcommand{\Xmark}{\textcolor{lightergray}{\ding{55}}}
\newcommand{\mycheck}{\textcolor{ForestGreen}{\ding{51}}}


\numberwithin{save@problem}{subsection}
\numberwithin{save@equation}{subsection}

\begin{document}
\title{DCFNet: Doppler Correction Filter Network \\ for Integrated Sensing and Communication \\ in Multi-User MIMO-OFDM Systems}
\author{Hyeonho Noh,~\IEEEmembership{Member,~IEEE}, Hyeonsu Lyu,~\IEEEmembership{Student Member,~IEEE}, \\ Moe Z. Win,~\IEEEmembership{Fellow,~IEEE}, and Hyun Jong Yang,~\IEEEmembership{Senior Member,~IEEE}
\thanks{Hyeonho Noh and Hyun Jong Yang are with the Department of Electrical and Computer Engineering, Seoul National University, Korea (email: \{hyeonho, hjyang\}@snu.ac.kr).
Hyeonsu Lyu is with the Department of Electrical Engineering, POSTECH, Korea (e-mail: hslyu4@postech.ac.kr). M. Z. Win is with the Laboratory for Information and Decision Systems, Massachusetts Institute of Technology, 77 Massachusetts Avenue, Cambridge, MA 02139, USA (e-mail: moewin@mit.edu).
}
}

% 통합 센싱 및 통신(ISAC)은 다가오는 IMT-2030 및 6G 표준의 핵심 기능으로 주목받고 있지만, 기존 직교 주파수 분할 다중 접속(OFDM) 체계를 활용한 complete ISAC 시스템을 제안한 구체적인 연구는 거의 진행되지 않았다. 대부분의 연구는 송신 빔포밍, 수신 신호 처리 등의 individual한 부분만 separately 최적화한다. 특히 OFDM을 이용한 센싱의 경우, 도플러로 인한 반송파 간 간섭(ICI)은 서브캐리어의 직교성을 파괴하여 거리-속도 맵을 흐리게 만들고 센싱 정확도를 심각하게 저하시킨다. 본 논문은 다중 사용자 다중 입력 다중 출력(MIMO) OFDM 시스템을 기반으로, 기존 프레임 구조를 변경하지 않으면서도 낮은 복잡도로 정밀한 거리-속도 해상도를 제공하는 송수신 빔포밍 최적화 기법과 AI 기반 ISAC 모델인 도플러 보정 필터 네트워크(DCFNet)를 제안한다. 송수신 최적화의 경우 
%  DCFNet은 먼저 여러 DCF 필터를 통해 주요 도플러 영역의 ICI 에너지를 우회시키고, 이후 경량 딥러닝 네트워크가 해당 간섭을 효과적으로 억제한다.

% 거리 및 속도 해상도를 더욱 향상시키기 위해, 본 논문은 일반화 우도비(GLRT)를 활용하여 DCFNet의 추정 결과를 서브셀 정확도로 정제하는 로컬 리파인먼트 기반 DCFNet(DCFNet-LR)을 추가로 제안한다. 시뮬레이션 결과, DCFNet-LR은 최대우도 탐색 방식 대비 약 143배 빠른 속도를 보이며, 다른 기존 기법들보다 일관되게 우수한 성능을 나타낸다.

\maketitle
\begin{abstract}\label{abstract}
Integrated sensing and communication (ISAC) is a headline feature for the forthcoming IMT-2030 and 6G releases, yet a concrete solution that fits within the established orthogonal frequency division multiplexing (OFDM) family remains open. Specifically, Doppler-induced inter-carrier interference (ICI) destroys sub-carrier orthogonality of OFDM sensing signals, blurring range-velocity maps and severely degrading sensing accuracy. Building on multi-user multi-input-multi-output (MIMO) OFDM systems, this paper proposes Doppler-Correction Filter Network (DCFNet), an AI-native ISAC model that delivers fine range-velocity resolution at minimal complexity without altering the legacy frame structure. A bank of DCFs first shifts dominant ICI energy away from critical Doppler bins; a compact deep learning network then suppresses the ICI. 
To further enhance the range and velocity resolutions, we propose DCFNet with local refinement (DCFNet-LR), which applies a generalized likelihood ratio test (GLRT) to refine target estimates of DCFNet to sub-cell accuracy.
Simulation results show that DCFNet-LR runs $143\times$ faster than maximum likelihood search and achieves significantly superior performance, reducing the range RMSE by up to $2.7 \times 10^{-4}$ times and the velocity RMSE by $6.7 \times 10^{-4}$ times compared to conventional detection methods.

% Integrated sensing and communication (ISAC) is a headline feature for the forthcoming IMT-2030 and 6G releases, yet a concrete solution that fits within the established orthogonal frequency division multiplexing (OFDM) family remains open. In high-mobility scenarios, Doppler-induced inter-carrier interference (ICI) destroys sub-carrier orthogonality, blurring range-velocity maps and severely degrading sensing accuracy. Building on multi-user multi-input-multi-output (MIMO) OFDM systems, this paper proposes Doppler-Correction Filter Network (DCFNet), an AI-native ISAC model that delivers fine range-velocity resolution at minimal complexity without altering the legacy frame structure. A bank of DCFs first shifts dominant ICI energy away from critical Doppler bins; a compact deep learning network then suppresses the ICI. 
% To further enhance the range and velocity resolutions, we propose DCFNet with local refinement (DCFNet-LR), which applies a generalized likelihood ratio test (GLRT) to refine target estimates of DCFNet to sub-cell accuracy.
% Simulation results show that DCFNet-LR runs $143\times$ faster than maximum-likelihood search, and consistently outperforms other baseline methods. 

\end{abstract}


\section{Introduction}
\label{sec:introduction}
In the forthcoming 6G era, integrated sensing and communication (ISAC) has emerged as a cornerstone technology underpinning next-generation wireless networks\cite{Wan24_VTM, Dong25_JSAC}. By unifying communication and sensing functionalities, ISAC empowers networks to not only deliver ultra-high-speed data but also perceive and interpret their environment in real-time\cite{Gonz24_proceeding, Dong25_JSAC}. This dual capability is pivotal for transformative applications including smart cities, autonomous driving, and immersive extended reality (AR/VR/XR), where future networks must simultaneously ensure reliable connectivity and continuously acquire rich contextual information\cite{Liu25_TWC, Wu25_JSAC}.

A key strategy for realizing ISAC is the design of dual-function radar-communication (DFRC) waveforms capable of concurrently performing radar sensing and wireless communication tasks\cite{Zhang24_TWC}. By enabling the reuse of spectral and hardware resources, DFRC systems promise significant improvements in spectral efficiency and cost savings by deploying a single integrated platform for both communication and radar sensing\cite{Wei22_CM}. However, there is a fundamental challenge in DFRC waveform design due to the inherent trade-off between the distinct requirements of sensing and communication\cite{Liu22_JSAC}. Sensing-centric waveforms, such as frequency modulated continuous wave (FMCW), offer large bandwidth and excellent autocorrelation properties, facilitating high-resolution target detection. Nevertheless, these waveforms typically fall short in supporting the sophisticated modulation schemes required for high-throughput communication. In contrast, communication-centric waveforms, exemplified by orthogonal frequency division multiplexing (OFDM), inherently accommodate high data rates and are already adopted in 5G and beyond standards\cite{Liu22_JSAC}. Additionally, OFDM naturally provides robust range resolution without suffering from range-Doppler coupling, making it particularly attractive for DFRC applications\cite{Xiao24_TSP}.

\begin{figure}[t]
    \centering
    \includegraphics[draft=false, width= \if 1\mycmd 0.7 \else 0.8 \fi \linewidth]{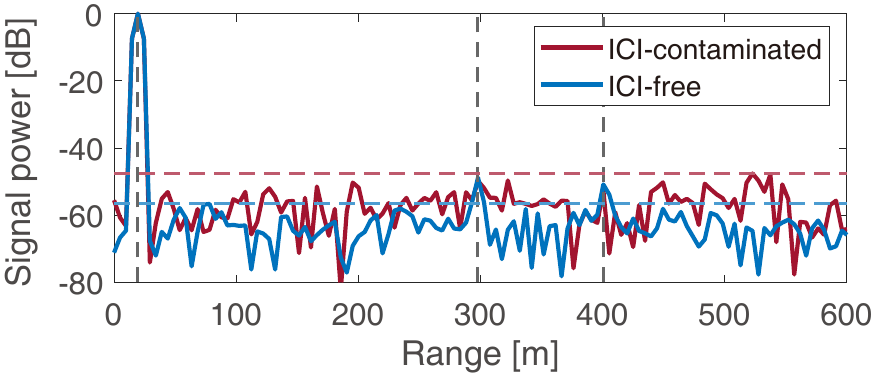}
    \vspace{-5pt}
    \caption{Normalized received signal power when the effect of ICI is included (ICI-contaminated) or ignored (ICI-free). Three targets are located on ($20~\text{m}$, $300~\text{m}$, and $400~\text{m}$). The velocities of all the targets are equivalent to $45~\text{m/s}$.}
    \label{fig:sidelobe_multiple_targets}
    \vspace{-5pt}
\end{figure}

% 순서를 1. Challenge 2. related works 3. Contribution 

% 1. 간단하게 related works의 문제점 지적 (+ table) 하고 AI의 개입이 필요함을 언급

% 3. contribution에서 end-to-end protocol에서 고려할 점을 모두 고려한 논문이다 언급 (transceiver beamforming, full stack physical layer protocol)

\subsection{Challenges}

A variety of approaches have been proposed for estimating target ranges and velocities in OFDM radar systems. These methods can be grouped into: (1) maximum likelihood (ML)-based methods \cite{Chabriel17, Mirabella23, Zhang20}, (2) subspace-based techniques \cite{Liu20, Xie21}, and (3) fast Fourier transform (FFT)-based solutions \cite{strum11, Tian17, Sit18, hakobyan18, Keskin21, Noh23}. ML-based methods can theoretically achieve near-optimal resolution but often rely on brute-force parameter searches, leading to substantial computational complexity. In contrast, subspace-based algorithms can achieve super-resolution performance without exhaustive parameter searches, yet they often require sufficiently high signal-to-noise ratio (SNR) and enough snapshots to reliably estimate the underlying signal subspace. Meanwhile, FFT-based techniques are conceptually straightforward and computationally efficient, though their resolution is fundamentally constrained by the OFDM waveform specifications. Moreover, as shown in Fig. \ref{fig:sidelobe_multiple_targets}, subspace- and FFT-based methods suffer from inter-carrier interference (ICI) at the radar receiver, which breaks subcarrier orthogonality, raises sidelobe levels, and ultimately degrades target detection performance \cite{Keskin21, Noh23}.

\begin{table}
\centering
    \caption{Comparison of the proposed scheme to recent works.
    }
    \vspace{-5pt}
    \begin{threeparttable}
    \adjustbox{width= \if 1\mycmd 0.6 \else 1.0 \fi \columnwidth}{
    \begin{tabular}{cccccc}
    \toprule
    \textbf{Ref} & \textbf{Scheme} & \textbf{Sens.} & \textbf{Comm.} & \textbf{Complexity} & \textbf{ICI suppr.} \\ \midrule[\heavyrulewidth]\midrule[\heavyrulewidth]
    \arrayrulecolor{lightgray}
    \cite{Mirabella23_TCOM, Zhang24_TWC, Xia25_IOTJ} & ML & \faThumbsOUp & \faThumbsOUp & \faThumbsDown & \Xmark \\ \cline{1-6}
    \cite{Zhang20} & ML & \faThumbsOUp \faThumbsOUp & \faThumbsOUp & \faThumbsDown & \mycheck \\ \cline{1-6}
    \cite{Xie21} & SS & \faThumbsOUp & \faThumbsOUp & \faThumbsDown & \Xmark \\ \cline{1-6}
    \cite{Hu24_TWC, Liu20} & SS & \faThumbsOUp & \faThumbsOUp & \faThumbsOUp & \Xmark \\ \cline{1-6}
    \cite{strum11, Xiao24_TSP, Liyanaarachchi24_TWC, Duan24_TVT} & FFT & \faThumbsDown & \faThumbsOUp & \faThumbsOUp \faThumbsOUp & \Xmark \\ \cline{1-6}
    \cite{Keskin21} & ML, FFT & \faThumbsOUp \faThumbsOUp & \faThumbsOUp & \faThumbsDown & \mycheck \\ \cline{1-6}
    \cite{Mirabella23_TCOM} & ML, FFT & \faThumbsOUp & \faThumbsOUp & \faThumbsOUp & \Xmark \\ \cline{1-6}
    \cite{hakobyan18} & FFT & \faThumbsOUp & \faThumbsDown & \faThumbsOUp \faThumbsOUp & \mycheck \\ \cline{1-6}
    \cite{Noh23, Park24_TVT} & FFT & \faThumbsOUp & \faThumbsOUp & \faThumbsOUp & \mycheck \\ \midrule
    % \midrule[\heavyrulewidth]
    \textbf{DCFNet} & \textbf{DL, ML, FFT} & \faThumbsOUp \faThumbsOUp & \faThumbsOUp & \faThumbsOUp & \mycheck \\ 
    \arrayrulecolor{black}
    \bottomrule
    % \\
    \end{tabular}
    }
    \vspace{0.02cm}
    \begin{flushleft}
    \footnotesize
    * SS: Subspace, Sens.: Sensing performance, 
    \\Comm.: Communication performance, ICI suppr.: ICI suppression.
    \end{flushleft}
    \end{threeparttable}
    \label{tab:compare}
    \vspace{-10pt}
\end{table}

Although previously developed analytical methods exhibit certain strengths, they also face inherent limitations, namely, the need for exhaustive computations, high SNR environments or susceptibility to performance degradation from ICI. These obstacles highlight that purely analytical approaches cannot fully address the complex trade-offs in OFDM-based ISAC systems. Motivated by these unresolved challenges, this study poses a key research question:
\begin{tcolorbox}[colframe=black, colback=white, height=1.5cm, boxrule=0.4mm]
\begin{center}
\vspace{-0.137cm}
\textit{\textbf{How can we develop an algorithm that delivers high-resolution sensing and maintain low computation cost while suppressing ICI?}}
\end{center}
\end{tcolorbox}

We adopt an advanced deep learning (DL) approach to effectively address this question, overcoming the inherent limitations in conventional analytic signal processing approaches.  
DL-driven solutions demonstrate unprecedented capabilities in capturing subtle patterns and mitigating complex interference effects, thereby enabling accurate, efficient, and robust target detection---an essential requirement for 6G wireless \cite{Wu24_CM}.

\subsection{Related Works}

\noindent\textbf{ML-based methods: }
ML-based methods can achieve higher accuracy but often involve considerable computational complexity.
Thus, most studies have focused heavily on reducing this complexity.
The work in \cite{Zhang20} proposed an alternating projection-maximum likelihood (AP-ML) method for joint range and Doppler estimation while suppressing ICI. Additionally, the expectation-maximization (EM) algorithm was employed to effectively reduce computational complexity.
The authors in \cite{Mirabella23} proposed an  algorithm that detects multiple targets using a serial cancellation and refinement procedure to estimate multiple radar parameters, achieving high resolution with lower complexity. The paper in \cite{Zhang24_TWC} proposes a unified tensor-based framework for joint channel and target parameter estimation in massive multi-input-multi-output (MIMO)-ISAC systems. The authors in \cite{Xia25_IOTJ} propose a semi-off-grid delay-Doppler estimation method for OFDM-ISAC systems.
% , using ambiguity-function-assisted refinement to achieve high accuracy with low complexity.

\noindent\textbf{Subspace methods: }
Subspace methods for joint range and velocity estimation include the well-known multiple signal classification (MUSIC) algorithm \cite{Schmidt86} and the estimation of signal parameters via rotational invariant technique (ESPRIT) \cite{Roy86}. 
In \cite{Xie21}, the authors presented the joint range-velocity estimator with 2D-MUSIC algorithm to overcome the inherent limitation in range and velocity resolution and investigate the performance of OFDM radar. 
However, MUSIC algorithm requires an exhaustive search over the entire parameter space to identify the most likely ranges, velocities, and angles of targets. To overcome the drawback of MUSIC algorithm, a novel auto-paired super-resolution range and velocity estimation method was proposed by enhancing the refinement of ESPRIT \cite{Liu20, Hu24_TWC}. By applying domain smoothing and exploiting translational invariance, it enables automatic pairing of range, velocity, and angle estimates without grid search.

% Although the existing researches have been evolved, subspace methods and ML-based methods are not still applicable in real-time application due to the prohibitive computational complexity. In the case of Fourier transform-based methods, the majorities of works ignore the impact of the ICI, which can be a critical problem when it comes to the high-mobility scenarios. 

\noindent\textbf{FFT-based methods: }
Sturm \textit{et al.} proposed a novel FFT-based OFDM radar system \cite{strum11, Tian17} with a phase-coded OFDM integrated waveform. The authors in \cite{Sit18} implemented real-time measurements to demonstrate the capability of target parameter estimation with universal software radio peripherals (USRPs).
To address the issue of ICI, some studies \cite{hakobyan18, Keskin21, Noh23} have proposed ICI mitigation techniques based on the Fourier transform-based method. The work in \cite{hakobyan18} utilized a symbol repetition-based ICI rejection method. However, the symbol repetition approach sacrifices the communication data rate significantly, which is not suitable for ISAC systems. The authors in \cite{Keskin21} proposed a successive interference cancellation-based method which first estimates the ICI caused by a specific target and utilizes the ICI-decontaminated radar range-velocity map (RV map) to estimate the range and velocity of that target. 
% However, the method proposed in \cite{Keskin21} requires a significant compute time due to two key factors. Firstly, the problem of estimating the ICI is inherently non-convex, which increases the complexity of the estimation process. Secondly, the iterative nature of the method poses additional computational demands, particularly when dealing with scenarios involving multiple targets. 
Noh \textit{et al.} \cite{Noh23} proposed an ICI-rejection method with the receiver beamformer optimization. 
% However, it also needs some iteration for solving the optimization problem. 
% Consequently, there is no method available that mitigates the impact of the ICI while maintaining a constant compute time irrespective of the number of targets.

\subsection{Contributions}
This paper establishes an end-to-end MIMO ISAC framework using OFDM radar integrated with multi-user (MU) communication. Specifically, we propose an innovative AI-driven radar sensing model, named Doppler Correction Filter Network (DCFNet), designed to mitigate ICI and efficiently estimate targets' ranges and velocities. 
The contributions are summarized as follows:
\begin{itemize}[leftmargin=*]
\item We introduce a complete MIMO-OFDM ISAC system that optimizes transceiver beamforming for MU communication and sensing-target detection. An optimization problem is formulated to maximize the sum-rate of users while guaranteeing sensing signal-to-noise-plus-interference ratio (SINR). By applying Lagrangian duality and the quadratic transform, we reformulate the non-convex formulation into a more tractable convex problem and solve it efficiently with an alternating optimization algorithm.
\item We propose DCFNet, a DL-based OFDM radar-sensing framework expressly designed to estimate targets' ranges and velocities from reflected signals. DCFNet comprises three key components: a bank of DCFs, an ICI rejection head, and a detection head. The DCF bank comprises multiple filters, each tailored to suppress ICI corresponding to a specific target velocity. The filtered radar images are then processed by the ICI rejection head, which generates interference-free latent representations. Finally, the detection head produces a confidence map indicating the likelihood of target presence at each range-velocity pixel. 
By mitigating Doppler-induced ICI in the received sensing signals, DCFNet enables accurate yet computationally efficient target detection.
\item To overcome the resolution ceiling inherent to FFT-based processing, we propose a DCFNet with local refinement (DCFNet-LR) which utilizes generalized likelihood-ratio test (GLRT). By leveraging the coarse range-velocity estimates from DCFNet, DCFNet-LR restricts the search to a localized set of candidate cells and refines the estimates with sub-cell precision. Consequently, DCFNet-LR achieves detection performance comparable to a full ML search over a finely discretized range-velocity grid, while substantially reducing computational complexity and avoiding the need for iterative estimation or optimization procedures.
\item Comprehensive simulations demonstrate that DCFNet offers substantial gains in both target detection and estimation performance, as well as computational efficiency, compared to conventional target detection methods. The performance closely approaches ideal, interference-free radar observations, validating DCFNet's effectiveness and highlighting its capability to robustly suppress ICI in practical OFDM-based ISAC scenarios.
\end{itemize}

\textit{Notations}: $(\cdot)^{*}$, $(\cdot)^\text{T}$, and $(\cdot)^\text{H}$ are complex
conjugate, transpose, and conjugate transpose, respectively. 
For a matrix $\mathbf{A}$, $\mathbf{A}^{-1}$, $|\mathbf{A}|$, and $\left[\mathbf{A}\right]_{(i,j)}$ are the inverse, absolute, and $(i,j)$-th entry of $\mathbf{A}$, respectively. The symbol $./$ is the elementwise complex division. The $N \times N$ identity matrix is denoted by $\mathbf{I}_N$. The diagonal matrix with diagonal elements $(a_1, a_2, \ldots,a_m)$ is denoted by $\text{diag}(a_1, a_2, \ldots, a_m)$. 
The complex normal distribution is denoted by $\mathcal{CN} (\mu, \Sigma)$ with mean $\mu$ and covariance matrix $\Sigma$. $\mathbb{R}^{n \times m}$ and $\mathbb{C}^{n \times m}$ are the $(n \times m)$-dimensional real and complex spaces, respectively. $\mathbb{E}(\cdot)$ is the expectation.  $\left\| \cdot \right\|_\text{F}$ is the Frobenius norm.

\begin{figure*}[t]
\centering
\includegraphics[draft=false, width=\if 1\mycmd 0.8 \else 1.0 \fi\textwidth]{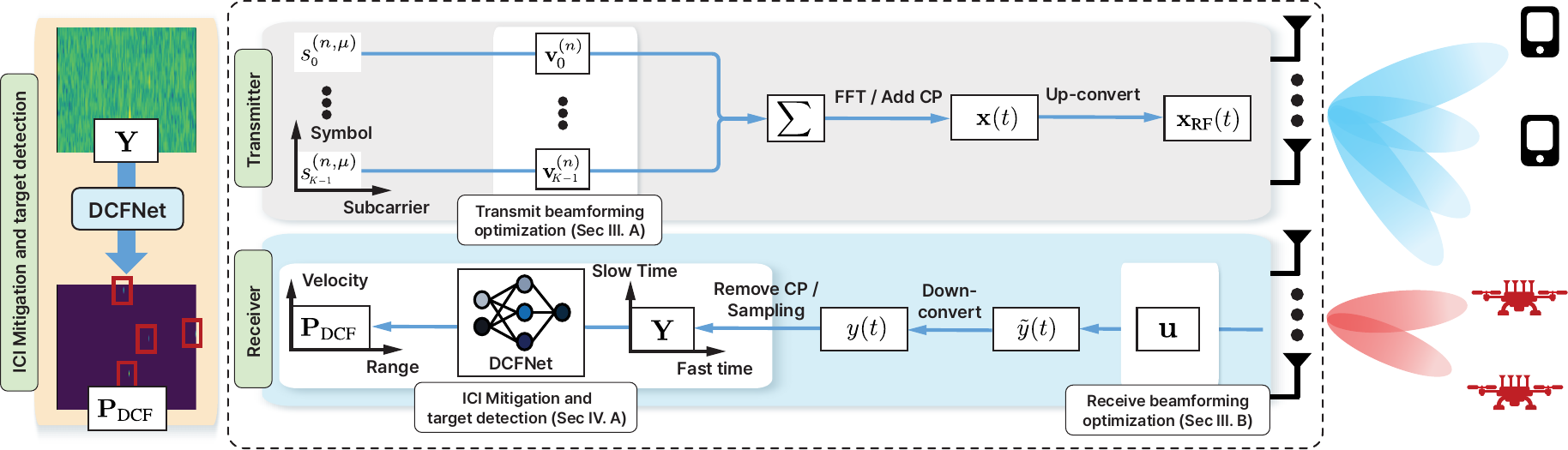}
\vspace{-6pt}
\caption{The proposed MU MIMO-OFDM ISAC system with the optimized transceiver and DCFNet for ICI mitigation and target detection.}
\label{fig:overall_process}
\vspace{-6pt}
\end{figure*}

\section{Scenario and Protocol}
\label{sec:system_model}

In Fig.~\ref{fig:overall_process}, we illustrate the proposed MU MIMO-OFDM ISAC system consisting of a base station (BS), $J$ radar targets, and $K$ communication users. The BS is equipped with a uniform linear array comprising $N_\text{T}$ transmit antennas and $N_\text{R}$ receive antennas, while each user is equipped with a single receive antenna.
The BS simultaneously serves the users and performs target sensing. Its primary objective is to estimate the ranges and velocities of targets while transmitting data to $K$ users. The ISAC system operates in two stages:

\noindent\textbf{ISAC signal transmission:} The BS transmits OFDM signals embedded with communication symbols, which are used both for downlink communication and target sensing within designated radar look directions. To improve the downlink data rate and enhance beamforming gain toward radar look directions, the BS applies a transmit beamforming strategy.

\noindent\textbf{Sensing signal reception:} The BS employs receive beamforming to amplify the signals reflected from the targets located in the radar look directions and suppress undesired multipath signals and clutters originating from other directions.

We assume far-field targets located along the line-of-sight (LOS) of the BS, moving at constant velocities. Each user may also be mobile, typically at walking speed. The communication environment is characterized by the presence of multiple scatterers surrounding the users, resulting in multipath propagation from the BS to each user.

\subsection{System and Signal Model}
For the duration of the consecutive $N_\textrm{sym}$ OFDM signals, the complex-valued time-domain OFDM signal vector at the baseband is written by
\begin{align} \label{def:OFDM_signal}
     \hspace{-0.20cm}\mathbf{x}(t) &=\frac{1}{\sqrt{N_\text{c}}} \sum_{\mu=0}^{N_\text{sym}-1} \sum_{n=0}^{N_\text{c}-1} \hspace{-2pt} \sum_{k \in \mathcal{K}} \mathbf{v}_{k}^{(n)} s_{k}^{(n,\mu)}  {e}^{j2\pi n\Delta f t_\mu} \nonumber \\[-.3cm]
     & \hspace{4.81cm} \cdot\text{rect} \Big( \frac{t_\mu+T_{\text{CP}}}{T_{\text{OFDM}}} \Big),
\end{align}
where the following variable definitions are used: $N_\text{c}$ is the number of OFDM subcarriers; $\mathbf{v}_{k}^{(n)} \in \mathbb{C}^{N_\text{T} \times 1}$ is the transmit beamforming vector designed for the $\mu$-th OFDM signal of the $k$-th user on the $n$-th subcarrier; $s_{k}^{(n,\mu)}$ is the complex-valued communication symbol of the $k$-th user's $\mu$-th OFDM signal on the $n$-th subcarrier;
$\Delta f$ = $1/T$ is the subcarrier spacing;
$T$ is the OFDM signal duration; $T_\text{CP}$ is the duration of the cyclic prefix (CP); $T_\text{OFDM} = T + T_\text{CP}$; $t_\mu = t - \mu T_\text{OFDM}$ denotes the fast-time variable of the data part of the $\mu$-th OFDM signal.
The function $\text{rect}(\cdot)$ is the rectangular function defined as
\begin{align}
    \text{rect} \Big( \frac{t}{T} \Big) =\begin{cases}
    1, & \text{for} ~ 0 \leq t < T,\\
    0, & \text{otherwise}.
  \end{cases}
\end{align}
It is assumed that channel coherence time is much longer than the total duration of OFDM signal transmission time $N_\text{sym} T_\text{OFDM}$; the transmit beamformer $\mathbf{v}_k^{(n)}$ remains constant over the channel coherence time. The BS transmits the up-converted OFDM signal as $\mathbf{x}_\text{RF}(t) = \mathbf{x}(t) e^{j 2 \pi f_\text{c} t}$.

\subsection{Communication SINR} 
\label{communication_SINR}
Each user receives $\mathbf{x}_\text{RF} (t)$ via a multi-path fading channel from the BS to the user. After that, the data part of the signal is obtained by detecting the end of the CP.
The users sample their received signals at the Nyquist sampling rate. After the FFT is applied to the time-domain sampled signal, the SINR of the $k$-th user on the $n$-th subcarrier is 
\begin{align}
    \gamma_k^{(n)} &= \frac{ \Big\|(\mathbf{h}_k^{(n)})^\text{H} \mathbf{v}_k^{(n)}\Big\|^2 }{\sum_{\ell \in \mathcal{K} \backslash \{ k\} } \Big\|(\mathbf{h}_k^{(n)})^\text{H} \mathbf{v}_\ell^{(n)}\Big\|^2 + N_\text{c} \sigma_k^2}, ~~ k \in \mathcal{K}.
\end{align}
where $\mathbf{h}_k^{(n)} \in \mathbb{C}^{
N_\text{T} \times 1}$ is the frequency-domain communication channel vector from the BS to the $k$-th user on the $n$-th subcarrier; $\sigma_k^2$ is the noise variance; $\mathcal{K} = \{0, 1, \ldots, K-1\}$.

\subsection{Radar Received Signal} 
The transmitted signal $\mathbf{x}_\text{RF}(t)$ is reflected by targets or clutters, and the reflected signals are received at the BS. The BS applies receive beamformer $\mathbf{u} \in \mathbb{C}^{N_\text{R} \times 1}$ to suppress clutter and extract target signals from the radar look direction.
% We start with the assumption that there exist targets in the radar look direction. 
% In a later section, we introduce a method for detecting the existence of a target in the radar look direction. 
% If there is no target in the current radar look direction, the radar look direction should be gradually changed to detect targets.  
In this work, we assume a constant target gain during $N_\text{sym} T_\text{OFDM}$ as in \cite{hakobyan18, Noh23}. 
Then, the received OFDM signal reflected by the targets and clutters is represented as 
\begin{align}\label{def:received_OFDM_signal}
    \hspace{-3.5pt} \tilde{y}(t) = \sum_{i \in \mathcal{P}} a_i \mathbf{u}^\text{H} \mathbf{A}_i \mathbf{x}(t-\tilde{\tau}_i(t))e^{j 2 \pi f_\text{c} \big(t-\tilde{\tau}_i(t)\big)} \hspace{-0.5pt} + \hspace{-0.5pt} \tilde{z}(t),
\end{align}
where the following variable definitions are used: $\mathcal{P}$ is the set of propagation paths, $\tilde{\tau}_i(t)$ is the time-dependent delay of the $i$-th path; $a_i$ is the complex-valued reflection coefficient of the $i$-th path; $\mathbf{A}_i$ is the phase shift matrix imposed on the $i$-th reflected  signal; $\tilde{z}(t)$ denotes the complex additive white Gaussian noise with distribution $\mathcal{CN}(0, \sigma^2)$.
With the far-field propagation assumption, $\mathbf{A}_i$ can be written by \cite{Stoica07} $\mathbf{A}_i = \mathbf{a}_\text{R}(\theta_{i}) \mathbf{a}_\text{T}^\text{H}(\phi_{i})$, where $\mathbf{a}_\text{M}(\theta) = \left[ 1, e^{-j 2 \pi \Delta \text{sin}(\theta)} , \ldots, e^{-j 2 \pi (N_\text{M}-1)\Delta \text{sin}(\theta)} \right]^\text{T} \in \mathbb{C}^{N_\text{M} \times 1}$ is the steering vector of the antenna array at $\theta$ for $\text{M} \in \{ \text{T}, \text{R} \}$, $\Delta$ denotes the antenna separation between adjacent antenna elements normalized by the wavelength \cite{Hassanien10, Hassanien16}, $\theta_i$ is the angle of arrival (AoA) of the $i$-th path, and $\phi_i$ is the angle of departure (AoD) of the $i$-th path.

The range to the $i$-th path is denoted by $R_i$ at $t=0$, and the target moves at a relative velocity of $v_i$. Thus, the time-dependent delay is given by $\tilde{\tau}_i(t) = 2(R_i + v_i t)/c_0.$
Then, down-converting the received RF signal in (\ref{def:received_OFDM_signal}) to the baseband, we get the baseband received signal at the BS as
\begin{align}\label{def:down_converted_signal}
    y(t) =& \tilde{y}(t) e^{-j 2 \pi f_\text{c} t} \nonumber \\ 
    =& \sum_{i \in \mathcal{P}} \sum_{\mu=0}^{N_\text{sym} - 1} \sum_{n=0}^{N_\text{c}-1} \sum_{k \in \mathcal{K}}  \frac{a_i}{\sqrt{N_\text{c}}} \mathbf{u}^\text{H} \mathbf{A}_i \mathbf{v}_{k}^{(n)} s_k^{(n,\mu)}  \nonumber \\ & \cdot e^{j2\pi n\Delta f (t_\mu- \tau_i)} e^{j 2 \pi f_{\text{D},i} t} e^{-j 2 \pi f_\text{c} \tau_i} e^{-j2\pi n \Delta f \frac{2v_i}{c_\text{0}}t} \nonumber \\ & \cdot \text{rect}\bigg(\frac{t_\mu + T_{\text{CP}} - \tau_i - \frac{2v_it}{c_\text{0}}}{T_{\text{OFDM}}}\bigg) + z(t),
\end{align}
where the Doppler shift is denoted by $f_{\text{D},i} = - 2v f_\text{c} / c_\text{0}$, the delay of the target at $t=0$ is rewritten as $\tau_i = 2R_i/c_\text{0}$, and $z(t) = \tilde{z}(t) e^{-j2\pi f_\text{c}t}$ is the baseband noise. 
Following \cite{hakobyan18, Noh23}, we assume that the range migration effect can be neglected, i.e., $e^{j2\pi n \Delta f \frac{2v}{c_\text{0}}t} \approx 1$.
Clearly, we get  $2vt/c_\text{0} T_\text{OFDM} \approx 0 $ in the rect-function.

Note that $y(t)$ is the received echo from the multiple propagation paths, which consists of a CP part and a data part. Provided that the CP length has been properly chosen to satisfy $\tau \leq T_\text{CP}$ and combat inter-symbol interference between the communication symbols, the receiver extracts the data part in the received signal by searching the end of the CP.
Then, for the $\mu$-th OFDM signal duration, the receiver samples the extracted data part at the Nyquist sampling rate; that is, $y(t)$ is sampled at $t_\mu = (m/N_\text{c}) T, ~m \in \mathcal{N}_\text{c} = \left\{0, 1, \ldots, N_\text{c}-1 \right\}$. The discretized received signal on the $m$-th sample during the $\mu$-th OFDM signal time is obtained as 
\begin{align}\label{extended_disc_radar}
    y^{(m,\mu)} \hspace{-2pt} & = y\big(\frac{m}{N_\text{c}}T + \mu T_\text{OFDM}\big) \nonumber \\
    & = \sum_{i \in \mathcal{P}} \sum_{n=0}^{N_\text{c}-1}  \sum_{k \in \mathcal{K}} \frac{\bar{a}_i}{\sqrt{N_\text{c}}} \mathbf{u}^\text{H} \mathbf{A}_i \mathbf{v}_{k}^{(n)} s_k^{(n,\mu)} e^{j2\pi n \frac{m}{N_c}} \nonumber \\
    & \hspace{10pt} \cdot e^{-j 2 \pi n \Bar{\tau}_i} e^{j 2 \pi \Bar{f}_{\text{D},i} \frac{m}{N_c}} e^{j 2 \pi \Bar{f}_{\text{D},i} \mu \alpha} + \hat{z}^{(m,\mu)},
\end{align}
where $\Bar{f}_{\text{D},i} = f_{\text{D},i}/\Delta f = f_{\text{D},i} T$ denotes the normalized Doppler shift, $\Bar{\tau}_i = \tau_i \Delta f = \tau_i / T$ denotes the normalized time delay, and $\hat{z}^{(m,\mu)} = \mathbf{u}^{\text{H}} \mathbf{z}^{(m,\mu)} \sim \mathcal{CN}(0, \sigma^2)$. In addition, $\bar{a}_i = a_i e^{-j 2 \pi f_\text{c} \tau_i}$ and $\alpha =T_\text{OFDM}/T $. 

The discretized received signal in (\ref{extended_disc_radar}) is rearranged to the matrix notation $\mathbf{Y}$, the rows and columns of which denote the fast-time and slow-time domains as follows:
\begin{align}\label{disc_received_signal_mat_form}
    \mathbf{Y} = \sum_{i \in \mathcal{P}} \bar{a}_i \mathbf{D}_{\text{I}}\big(\bar{f}_{\text{D},i}\big) \mathbf{F}_{N_\text{c}}^{-1} \mathbf{D}_{\text{R}}^*\big(\bar{\tau}_{i}\big) \mathbf{S}_i \mathbf{D}_{\text{v}}\big(\bar{f}_{\text{D},i}\big) + \mathbf{Z}.
\end{align}
Here, $\mathbf{S}$, 
$\mathbf{D}_{\text{I}}(f), \mathbf{D}_{\text{R}}(\tau)$, and $\mathbf{D}_{\text{v}}(f)$ are defined by 
\begin{align}
    [\mathbf{S}_i]_{(p,q)} & = \mathbf{u}^\text{H} \mathbf{A}_i \sum\limits_{k \in \mathcal{K}} \mathbf{v}_k^{(p)} s_k^{(p,q)} \label{subeq:s_def}, \\ 
    \mathbf{D}_{\text{I}}(f) & = \text{diag} \left( 1, e^{j2\pi \frac{f}{N_\text{c}}}, \ldots, e^{j2\pi \frac{f}{N_\text{c}} (N_\text{c}-1)} \right), \\
    \mathbf{D}_{\text{R}}(\tau) & = \text{diag} \left( 1, e^{j2\pi \tau}, \ldots, e^{j2\pi \tau (N_\text{c}-1)} \right), \\ 
    \mathbf{D}_{\text{v}}(f) & = \text{diag} \left( 1, e^{j2\pi f \alpha}, \ldots, e^{j2\pi f \alpha (N_\text{sym}-1)} \right),
\end{align}
the $N$-point discrete Fourier transform matrix $\mathbf{F}_{N}$ is \begin{align}
    \left[\mathbf{F}_{N}\right]_{(p,q)} = \frac{1}{\sqrt{N}} e^{-j \frac{2 \pi}{N}pq}, ~~~ 0 \le p,q \le N-1,
\end{align}
and $\left[\mathbf{Z}\right]_{(n,\mu)} = \hat{z}^{(n,\mu)}$ is defined as the $(n,\mu)$-th element of $\mathbf{Z}$.

In the process of estimating target ranges and velocities, the ICI term $\mathbf{D}_{\text{I}}(\bar{f}_{\text{D}, i})$ induces frequency-domain misalignment, which severely degrades detection performance. To mitigate the impact of $\mathbf{D}_{\text{I}}$, we propose a DL-driven sensing model, named DCFNet, as detailed in Sec.~\ref{sec:proposed_scheme}.

% \remark
% In estimating the range and velocity of the targets, the ICI term $\mathbf{D}_{\text{I}}(\bar{f}_{\text{D}, i})$ causes misalignment in the frequency domain, which significantly deteriorates the target detection performance.
% In the early stages of OFDM radar research  \cite{strum11}, target detection was performed by ignoring such ICI terms and applying conventional FFTs to both the fast-time and slow-time dimensions. However, this approach significantly impairs target detection performance when targets have high velocities or there are multiple targets.
% Following \cite{strum11}, the work in \cite{hakobyan18} proposed ACDC, by which the same OFDM symbols are transmitted repeatedly. Though ACDC removes the ICI effect clearly, it significantly sacrifices the communication data rate. \textit{Noh} et al. proposed the optimization-based transceiver design that rejects ICI while not sacrificing the data rate. However, the optimization approach can lead to considerable computation complexity. 
% In the subsequent section, we propose a deep learning-based scheme that rejects ICI.

\subsection{Radar FFT Process}
\label{subsec:2D FFT process for target detection}

\begin{figure*}[t]
\centering
\includegraphics[draft=false, width=\if 1\mycmd 0.8 \else 0.9 \fi\textwidth]{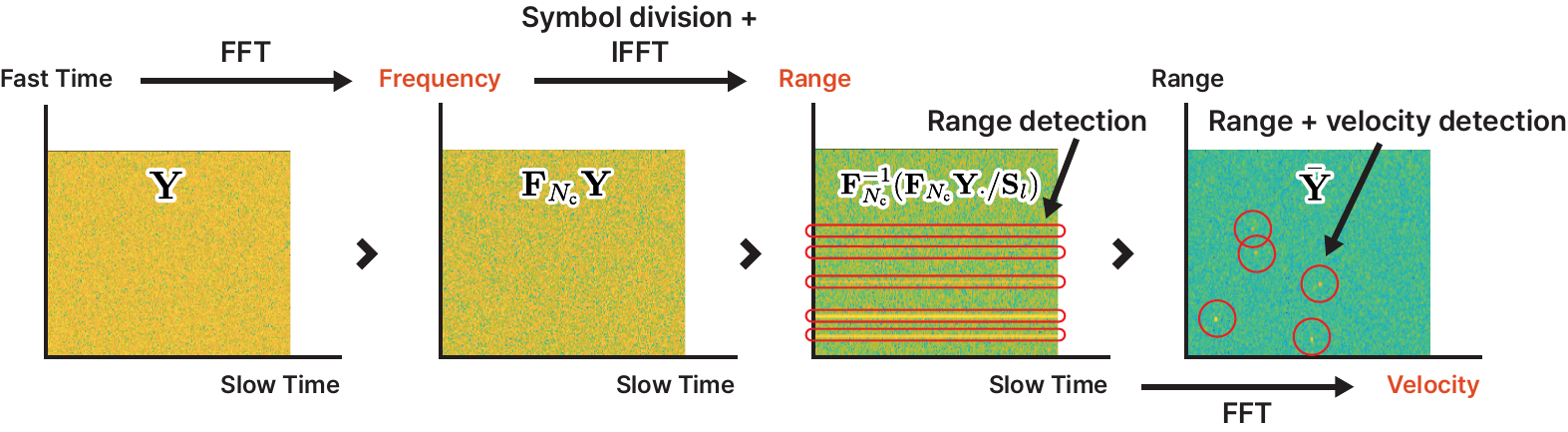}
\vspace{-0.4cm}
\caption{Range and velocity estimation pipeline in ISAC receiver.}
\label{fig:radar_image_processing}
\vspace{-6pt}
\end{figure*}

Although FFT-based method is widely used for estimating target ranges and velocities, most prior works neglect the impact of ICI from $\mathbf{D}_\text{I}(\bar{f}_{\text{D}, i})$ for tractability. However, this assumption over-simplifies the problem, so it is not applicable when Doppler shifts are non-negligible. We derive here the FFT-based target detection method without such simplification.

The overall FFT process is depicted in Fig. \ref{fig:radar_image_processing}. We define the set of propagation paths as $\mathcal{P}_{\theta_l}$, which encompasses all propagation path indices corresponding to the focal angle $\theta_l$ of the BS. In this case, for any $i \in \mathcal{P}_l$, $\mathbf{S}_l \approx \mathbf{S}_i$ holds. Then, the discretized received signal with matrix form in \eqref{disc_received_signal_mat_form} reads 
\begin{align}\label{disc_received_signal_mat_form_divide}
    \mathbf{Y} &= \underbrace{\sum_{i \in \mathcal{P}_l} \bar{a}_i \mathbf{F}_{N_\text{c}}^{-1} \mathbf{D}_{\text{R}}^*\left(\bar{\tau}_{i}\right) \mathbf{S}_l \mathbf{D}_{\text{v}}\left(\bar{f}_{\text{D},i}\right)}_\text{Desired signal} 
    \\ &+ 
    \underbrace{\sum_{i \in \mathcal{P} \backslash \mathcal{P}_l} \bar{a}_i \mathbf{F}_{N_\text{c}}^{-1} \mathbf{D}_{\text{R}}^*\left(\bar{\tau}_{i}\right) \mathbf{S}_i \mathbf{D}_{\text{v}}\left(\bar{f}_{\text{D},i}\right)}_\text{Undesired multipath signal $\big(\!= \sum_{i \in \mathcal{P} \backslash \mathcal{P}_l} \mathbf{Y}_{\text{M},i}\big)$} \nonumber \\ &+ 
    \underbrace{\sum_{i \in \mathcal{P}} \bar{a}_i (\mathbf{D}_{\text{I}}\left(\bar{f}_{\text{D},i}\right)-\mathbf{I}_{N_\text{c}}) \mathbf{F}_{N_\text{c}}^{-1} \mathbf{D}_{\text{R}}^*\left(\bar{\tau}_{i}\right) \mathbf{S}_i \mathbf{D}_{\text{v}}\left(\bar{f}_{\text{D},i}\right)}_\text{Inter-carrier interference signal $\big(\!= \sum_{i \in \mathcal{P}}\mathbf{Y}_{\text{ICI},i}\big)$}
    + \mathbf{Z}. \nonumber
\end{align}
The first term in \eqref{disc_received_signal_mat_form_divide} represents the desired signal, i.e., the ICI-free component. The second term accounts for undesired multipath reflections from angles outside the focal direction $\theta_l$. The third term induces ICI, contaminating the desired signal. We denote the undesired multipath and ICI components as $\sum_{i \in \mathcal{P} \backslash \mathcal{P}_l} \mathbf{Y}_{\text{M},i}$ and $\sum_{i \in \mathcal{P}} \mathbf{Y}_{\text{ICI}, i}$, respectively.
Notably, $\mathbf{Y}_{\text{ICI}, i}$ vanishes as the velocity of the $i$-th target approaches zero.

% Unlike conventional radar systems, OFDM radar employs the transmission of communication symbol-combined signals. This unique approach causes the received signal to be modulated by the communication symbols. To accurately extract the radar parameters of the targets from the signals modulated by communication symbols, it is crucial to eliminate the influence of the symbols. 
Through the conventional FFT-based OFDM radar detection methods, we can obtain the RV map as
\begin{align}\label{eq:radar_image}
    \Bar{\mathbf{Y}} & = \mathbf{F}_{N_\text{c}}^{-1}  (\mathbf{F}_{N_\text{c}} \mathbf{Y} ./ \mathbf{S}_l) \mathbf{F}_{N_\text{sym}} \\ &= \sum_{i \in \mathcal{P}_l} \bar{a}_i \mathbf{e}^{*}_{\text{R}}(\bar{\tau}_i) \mathbf{e}_{\text{v}}^\text{T}\left(\bar{f}_{\text{D},i}\right) + \sum_{i \in \mathcal{P} \backslash \mathcal{P}_l } \Bar{\mathbf{Y}}_{\text{M}, i} + \sum_{i \in \mathcal{P}} \Bar{\mathbf{Y}}_{\text{ICI}, i} + \bar{\mathbf{Z}},\nonumber
\end{align}
where $\Bar{\mathbf{Y}}_{\text{M}, i} = \mathbf{F}_{N_\text{c}}^{-1}  (\mathbf{F}_{N_\text{c}} \mathbf{Y}_{\text{M}, i} ./ \mathbf{S}_l) \mathbf{F}_{N_\text{sym}}$, $\Bar{\mathbf{Y}}_{\text{ICI}, i} = \mathbf{F}_{N_\text{c}}^{-1} (\mathbf{F}_{N_\text{c}} \mathbf{Y}_{\text{ICI}, i} ./ \mathbf{S}_l) \mathbf{F}_{N_\text{sym}}$,  $\bar{\mathbf{Z}} = \mathbf{F}_{N_\text{c}}^{-1} (\mathbf{F}_{N_\text{c}} \mathbf{Z} ./ \mathbf{S}_l) \mathbf{F}_{N_\text{sym}}$, and $\mathbf{e}_{\text{R}}\in \mathbb{C}^{N_\text{c} \times 1}$ and $\mathbf{e}_{\text{v}}\in \mathbb{C}^{N_\text{sym} \times 1}$ are defined as
\begin{align} \label{eq:er}
    \mathbf{e}_\text{R}(\bar{\tau}) &= \frac{1}{\sqrt{N_\text{c}}} 
    \begin{pmatrix}
    \sum\limits_{n=0}^{N_\text{c}-1} e^{j 2 \pi \left(\bar{\tau} - \frac{0}{N_\text{c}}\right)n} \\ \vdots \\ \sum\limits_{n=0}^{N_\text{c}-1} e^{j 2 \pi \left(\bar{\tau} - \frac{N_\text{c}-1}{N_\text{c}}\right)n}
    \end{pmatrix},
\end{align}
\begin{align}
    \label{eq:ev} 
    \mathbf{e}_\text{v}(\bar{f}_\text{D}) &= \frac{1}{\sqrt{N_\text{sym}}}    
    \begin{pmatrix}
    \sum\limits_{n=0}^{N_\text{sym}-1} e^{j 2 \pi \left(\bar{f}_{\text{D}} \alpha - \frac{0}{N_\text{sym}}\right)n} \\ \vdots \\ \sum\limits_{n=0}^{N_\text{sym}-1} e^{j 2 \pi \left(\bar{f}_{\text{D}} \alpha - \frac{N_\text{sym}-1}{N_\text{sym}}\right)n}
    \end{pmatrix}.
\end{align}
In ideal ICI-free scenarios, the range and velocity of the target can be found by finding the peak points of the RV map $\left| \Bar{\mathbf{Y}} \right|$. However, the ICI term $\sum_{i \in \mathcal{P}} \Bar{\mathbf{Y}}_{\text{ICI}, i}$ contaminates the RV map by elevating sidelobe levels, potentially causing target misdetection.
Note that the conventional FFT-based radar detection process in \eqref{disc_received_signal_mat_form_divide} and \eqref{eq:radar_image} can also be interpreted as a functional operator $\Pi$ that operates on the input $\mathbf{Y}$ to produce the output $|\bar{\mathbf{Y}}|$, i.e., $\Pi(\mathbf{Y}): \mathbf{Y}  \xrightarrow{\Pi} |\bar{\mathbf{Y}}|$.

\section{Transceiver Beamforming Design}
\label{sec:beamforming_design}
\begin{figure}[t]
\centering
\includegraphics[draft=false, width=\if 1\mycmd 0.4 \else 0.95 \fi\columnwidth]{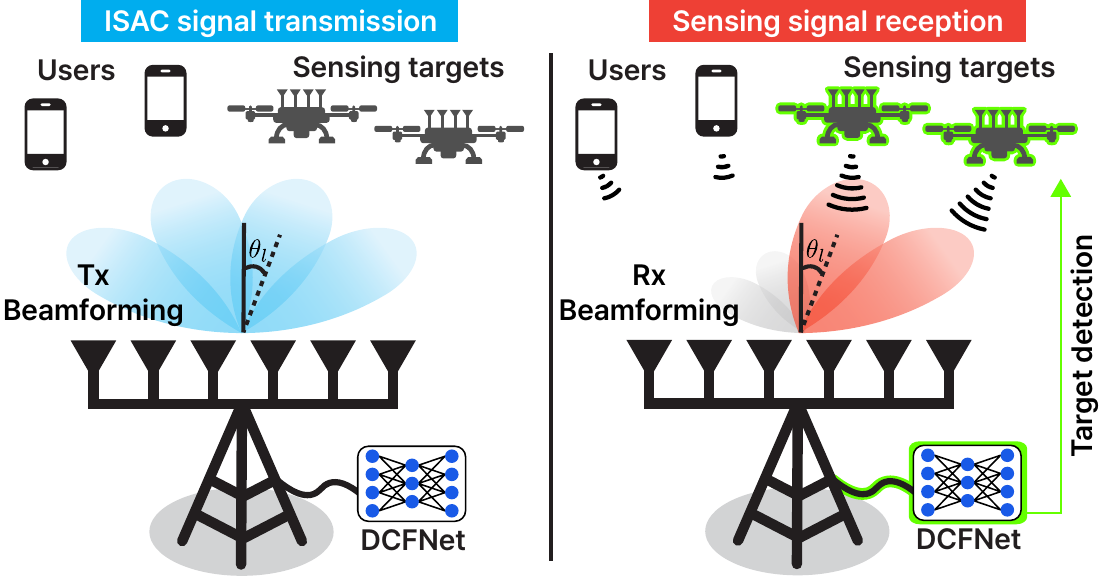}
\vspace{-6pt}
\caption{The proposed MU MIMO-OFDM ISAC transceiver beamforming.}
\label{fig:system_model}
\vspace{-6pt}
\end{figure}

The design of transmit and receive beamformers in the proposed ISAC system aims to jointly satisfy the requirements of both sensing and communication functionalities, as shown in Fig. \ref{fig:system_model}.
The transmit beamformer is designed to achieve the following objectives:
\begin{itemize}[leftmargin=*]
    \item Ensure that the sensing SINR in the configured radar look directions meets a predefined minimum threshold.
    \item Maximize the downlink sum-rate across all users.
    \item Satisfy the total transmit power limit at the BS.
\end{itemize}
To strike a balance between sensing accuracy and communication performance, the beamforming weights are carefully optimized such that sufficient energy is directed toward the radar look directions while maintaining high spectral efficiency for communication.

The receive beamformer is designed to enhance the detection capability of the BS with the following objectives:
\begin{itemize}[leftmargin=*]
    \item Maximize beamforming gain toward designated radar directions to enhance the reception of target reflections.
    \item Suppress undesired multipath signals and clutter returns arriving from directions outside the radar look directions.
\end{itemize}
This dual-purpose design facilitates robust target parameter estimation, forming the core of the ISAC receiver strategy.

\subsection{Transmit Beamformer Design}
\noindent\textbf{Problem formulation: } Based on (\ref{def:OFDM_signal}), the $m$-th sample of the baseband OFDM signal at $t_\mu = (m/N_c) T,~m \in \mathcal{N}_\text{c}$, on subcarrier $n$ during OFDM signal time $\mu$ is denoted as
\begin{align}
    \mathbf{x}_\text{OFDM}^{(n,\mu)}(m) = \frac{1}{\sqrt{N_\text{c}}}  \sum_{k=1}^K \mathbf{v}_k^{(n)} s_k^{(n,\mu)} e^{j2\pi n \frac{m}{N_\text{c}}}.
\end{align}
Assume that $s_k^{(n,\mu)}$ is independent and identically distributed with zero mean and unit variance for all $k \in \mathcal{K}$. Then, the beampattern gain at focal angle $\phi$ on subcarrier $n$ is given by
\begin{align}
    P_\text{T}^{(n)}(\phi) & = \mathbf{a}^\text{H} (\phi) \cdot \mathbb{E}_{s_k^{(n,\mu)}}\Big\{ \mathbf{x}_\text{OFDM}^{(n,\mu)}(m) \left(\mathbf{x}_\text{OFDM}^{(n,\mu)}(m)\right)^\text{H} \Big\} \mathbf{a} (\phi) \nonumber \\ 
    & = \frac{1}{N_\text{c}} \sum_{k \in \mathcal{K}} \mathbf{a}^\text{H} (\phi) \mathbf{v}_k^{(n)} \big(\mathbf{v}_k^{(n)}\big)^\text{H} \mathbf{a} (\phi).
\end{align}

By denoting the sum-rate as $R_\text{sum}^{(n)} = \sum_{k \in \mathcal{K}} \log (1 + \gamma_k^{(n)})$, the sum-rate maximization problem under the transmit power and minimum sensing SINR constraints can be formulated as
\begin{subequations}
\vspace{-0.3cm}
\label{eq:opt}
\begin{align}
    & \max_{ \{ \mathbf{v}_k^{(n)} \}} & & R_\text{sum}^{(n)}, \label{subeq:opt_a} \\ 
    & ~~\text{s.t.} & & \text{tr} \big( \sum_{k\in\mathcal{K}} \mathbf{v}_k^{(n)} (\mathbf{v}_k^{(n)})^\text{H} \big) \leq P_\text{BS}^{(n)}, \label{subeq:opt_b}\\ 
    & & & P_\text{T}^{(n)}(\phi_l) \geq P_{\text{T}, \text{req}}^{(n)}, l=1,\ldots, L, \label{subeq:opt_c}
\end{align}
\end{subequations}
where $P_{\text{T}, \text{req}}^{(n)}$ and $P_\text{BS}^{(n)}$ are the desired minimum beampattern gain toward the focal angles and power budget of the BS on subcarrier $n$, respectively, and $L$ is the number of focal angles.

Introducing $\mathbf{V}_k^{(n)} \triangleq \mathbf{v}_k^{(n)}(\mathbf{v}_k^{(n)})^{\mathrm H}$ allows problem \eqref{eq:opt} to be reformulated as a semidefinite program (SDP). Conventional works applied standard semidefinite relaxation (SDR) by dropping the non-convex rank-one constraints yields a convex problem solvable by tools such as CVX\cite{Hua23_TVT, Demirhan24_TCOM}. The relaxed solution, however, is generally not rank-one, and post-processing via eigenvalue decomposition or Gaussian randomization often violates the strict power constraint in \eqref{subeq:opt_c}. As a result, the synthesized beamformers may fail to exploit the available power budget, leading to sub-optimal sum-rate performance. Thus, we solve \eqref{eq:opt} via reformulation and alternating optimization.
\vspace{-0.2cm}

\subsection{Transmit Beamformer Design}
\label{subsec:tx_beamformer_design}
The optimization problem in \eqref{eq:opt} is non-concave as the objective function is non-concave. Therefore, we reformulate the objective into a more tractable form as follows.

\noindent\textbf{Lagrangian dual transform: } By introducing an auxiliary variable $\beta_k^{(n)}$, the objective function can be transformed into
\begin{align}
    R_\text{sum}^{(n)} = \hspace{-3pt} \max_{\beta_k^{(n)} \geq 0} ~\log\big(1+\beta_k^{(n)}\big) - \beta_k^{(n)} \hspace{-2pt} + \hspace{-2pt} \frac{(1 + \beta_k^{(n)}) \gamma_k^{(n)}}{1+\gamma_k^{(n)}}.
\end{align}
With the other parameters fixed, we have the optimal $\hat{\beta}_k^{(n)}$, which can be expressed as $\hat{\beta}_k^{(n)} = \gamma_k^{(n)}$. 

\noindent\textbf{Quadratic transform: }
Now we fix $\{\beta_k^{(n)}\}, \mathbf{u}$ and solve the optimization problem \eqref{eq:opt}. We adopt a quadratic transform to make the fractional form in $R_\text{sum}$ tractable, which is
\begin{align}
    \frac{(1+\beta_k^{(n)})\gamma_k^{(n)}}{1+\gamma_k^{(n)}} &= \frac{\big(1+\beta_k^{(n)}\big) \big\| (\mathbf{h}_k^{(n)})^\text{H} \mathbf{v}_k^{(n)}\big \|^2 }{\sum_{p \in \mathcal{K}}\big \| (\mathbf{h}_k^{(n)})^\text{H} \mathbf{v}_p^{(n)} \big\|^2 + N_\text{c} \sigma_k^2 }  \\
    &= \max_{\xi_k^{(n)}} \hspace{2pt} ~2\sqrt{1+\beta_k^{(n)}} \text{Re}(\xi_k^{(n)} (\mathbf{h}_k^{(n)})^\text{H} \mathbf{v}_k^{(n)})
    \nonumber \\
    &- |\xi_k^{(n)}|^2 \Big(\sum_{p \in \mathcal{K}} \big\| (\mathbf{h}_k^{(n)})^\text{H} \mathbf{v}_\ell^{(n)} \big\|^2 + N_\text{c} \sigma_k^2 \Big). \nonumber
\end{align}
Similar to $\beta_k^{(n)}$, with the other parameters fixed, we have the optimal $\xi_k^{(n)}$, which can be expressed as 
\begin{align}
    \hat{\xi}_k^{(n)} = \frac{\sqrt{1+\beta_k^{(n)}} (\mathbf{h}_k^{(n)})^\text{H} \mathbf{v}_k^{(n)}}{\sum_{p \in \mathcal{K}} \| (\mathbf{h}_k^{(n)})^\text{H} \mathbf{v}_p^{(n)} \|^2 + N_\text{c} \sigma_k^2}.
\end{align} 

\noindent\textbf{Reformulation: } The reformulated problem can be written as
\begin{subequations}
\label{eq:opt_reform}
\begin{align}
    & \max_{ \{ \mathbf{v}_k^{(n)} \}, \boldsymbol{\xi}, \boldsymbol{\beta} } & & \!\!\! \sum_{k \in \mathcal{K} } \Big(2\sqrt{1+\beta_k^{(n)}} \text{Re}(\xi_k^{(n)} (\mathbf{h}_k^{(n)})^\text{H} \mathbf{v}_k^{(n)}\Big)
    \nonumber \\
    & & & \!\!\! - |\xi_k^{(n)}|^2 \Big(\sum_{p \in \mathcal{K}} \| (\mathbf{h}_k^{(n)})^\text{H} \mathbf{v}_\ell^{(n)} \|^2 + N_\text{c} \sigma_k^2 )\Big), \label{subeq:opt_reform_a}\\
    & \text{s.t.} & & \eqref{subeq:opt_b}, \eqref{subeq:opt_c},
\end{align}
\end{subequations}
where $\boldsymbol{\xi} = \{ \xi_k^{(n)} \}_{n=0,\ldots,N_\text{c}-1, k \in \mathcal{K}}$ and $\boldsymbol{\beta} = \{ \beta_k^{(n)} \}_{n=0,\ldots,N_\text{c}-1, k \in \mathcal{K}}$. The optimization problem is a convex quadratic optimization problem in terms of $\mathbf{v}_k^{(n)}$ for given $\boldsymbol{\xi}$ and $\boldsymbol{\beta}$, which can be solved by using standard convex optimization algorithms. We obtain a closed-form solution of the optimal $\mathbf{v}_k^{(n)}$ by using the Lagrange multipliers method. 
By introducing the Lagrange multipliers $\lambda$ and $\{\mu_l\}$ for \eqref{subeq:opt_b} and \eqref{subeq:opt_c}, respectively, the first-order optimality condition yields the closed-form expression for $\mathbf{v}_k^{(n)}$:
\begin{align}
\label{w_k_lagrange}
    \mathbf{v}_{k}^{(n)}= \xi_k \sqrt{1+\beta_k} \bigg(\sum_{p \in \mathcal{K}} | & \xi_p |^2  \mathbf{h}_p^{(n)} (\mathbf{h}_p^{(n)})^\text{H} + \lambda \mathbf{I} \\[-0.3cm] 
    & - \frac{1}{N_\text{c}} \sum_{l=1}^L \mu_\ell \mathbf{a}(\phi_l) \mathbf{a}^\text{H}(\phi_l)\bigg)^{-1} \mathbf{h}_k^{(n)}.  \nonumber
\end{align}
% This expression corresponds to the solution obtained by differentiating the Lagrangian with respect to $\mathbf{v}_k^{(n)}$ and applying the KKT conditions.

% Attaching a Lagrange multiplier $\lambda$ and $\{\mu_l\}$ to the power and beampattern constraints, respectively, we define the Lagrange function \eqref{eq:lagrange} at the bottom of this page.
% \begin{table*}[!b]
% \hrule
% \begin{align}
% \label{eq:lagrange}
% \medskip
%     L( \{ \mathbf{v}_k\}, \lambda, \{\mu_l\} ) &=  \sum_{k \in \mathcal{K} } (2\sqrt{1+\beta_k} \text{Re}(\xi_k^\text{H} (\mathbf{h}_k^{(n)})^\text{H} \mathbf{v}_k^{(n)}) - | \xi_k |^2 (\sum_{\ell \in \mathcal{K}} \| (\mathbf{h}_k^{(n)})^\text{H} \mathbf{v}_\ell^{(n)} \|^2 + N_\text{c} \sigma_k^2 )) \nonumber \\ 
%     &- \lambda ( \text{tr} \left( \sum_{k\in\mathcal{K}} \mathbf{v}_k \mathbf{v}_k^\text{H} \right) - P_\text{BS}^{(n)}) - \mu_l ( \frac{1}{N_\text{c}} \sum_{l=1}^{L}  \sum_{k \in \mathcal{K}} \mathbf{a}^\text{H} (\phi_l) \mathbf{v}_k^{(n)} \left(\mathbf{v}_k^{(n)}\right)^\text{H} \mathbf{a} (\phi_l) ).
% \protect
% \end{align}
% \end{table*}
% The first-order optimality condition of the Lagrange function $L(\mathbf{v}^{(n)}, \lambda, \{ \mu_l\})$, with respect to $\mathbf{v}^{(n)}$, yields 
% \begin{align}
%     \mathbf{v}_{k}^{(n)}= \xi_k \sqrt{1+\beta_k} (&\sum_{p \in \mathcal{K}} | \xi_p |^2 \mathbf{h}_p^{(n)} (\mathbf{h}_p^{(n)})^\text{H} + \lambda \mathbf{I} \nonumber \\ &- \frac{1}{N_\text{c}} \sum_{l=1}^L \mu_\ell \mathbf{a}(\phi_l) \mathbf{a}^\text{H}(\phi_l))^{-1} \mathbf{h}_k^{(n)}, \label{w_k_lagrange}
% \end{align}

By the complementary slackness condition in terms of constraints \eqref{subeq:opt_b} and \eqref{subeq:opt_c}, we have two cases according to the value of the Lagrange multipliers, i.e., $\lambda$ and $\{ \mu_l \}$.

\subsubsection{Solution for $\lambda$} With the fixed $\boldsymbol{\mu} = \{ \mu_1, \ldots, \mu_L\}$, we let $\mathbf{v}_k^{(n)} (\lambda, \{\mu_l\})$ denote the right-hand side of \eqref{w_k_lagrange}. If $\sum_{p \in \mathcal{K}} | \xi_k |^2 \mathbf{h}_p^{(n)} (\mathbf{h}_p^{(n)})^\text{H} - \frac{1}{N_\text{c}} \sum_{l=1}^L \mu_\ell \mathbf{a}(\phi_l) \mathbf{a}^\text{H}(\phi_l)$ is invertible and $\text{tr} ( \sum_{k\in\mathcal{K}} \mathbf{v}_k^{(n)}(0, \boldsymbol{\mu}) (\mathbf{v}_k^{(n)} (0, \boldsymbol{\mu}))^\text{H} ) \leq P_\text{BS}^{(n)}$, we have $\hat{\lambda} = 0$. Otherwise, the constraint \eqref{subeq:opt_b} must hold with equality. Let $\mathbf{E} \boldsymbol{\Lambda} \mathbf{E}^\text{H}$ represent the eigenvalue decomposition of $\sum_{p \in \mathcal{K}} | \xi_p |^2 \mathbf{h}_p^{(n)} (\mathbf{h}_p^{(n)})^\text{H} - \frac{1}{N_\text{c}} \sum_{l=1}^L \mu_\ell \mathbf{a}(\phi_l) \mathbf{a}^\text{H}(\phi_l)$. Then, the following equation should be satisfied
\begin{align}
\label{eq:eigenvalue_decom}
    \sum_{k \in \mathcal{K} } \text{Tr} \big( (\boldsymbol{\Lambda} + \lambda \mathbf{I} )^{-2} \boldsymbol{\Upsilon}_k\big) = P_\text{BS}^{(n)},
\end{align}
where $\boldsymbol{\Upsilon}_k = |\xi_k^{(n)}|^2 (1+\beta_k^{(n)}) \mathbf{E}^\text{H} \mathbf{h}_k \mathbf{h}_k^\text{H} \mathbf{E}$. Then, the equation \eqref{eq:eigenvalue_decom} is equivalent to
\begin{align}
    \sum_{k \in \mathcal{K} } \sum_{i=1}^{N} \frac{[\boldsymbol{\Upsilon}_k]_{ii}}{( [\boldsymbol{\Lambda}]_{ii} + \lambda )^2} = P_\text{BS}^{(n)}.
\end{align}
With the fixed $\boldsymbol{\mu}$, the optimal value of $\lambda$ can be readily obtained via one-dimensional bisection methods.

\subsubsection{Solution for $\boldsymbol{\mu}$} 
The Lagrange multiplier $\mu_l$ is an implicit function of $\mathbf{v}_k^{(n)}$, making the closed-form solution intractable. Instead, we adopt an iterative optimization approach, optimizing $\mu_l$ while fixing $\mu_1, \ldots, \mu_{l-1}, \mu_{l+1}, \ldots, \mu_L$. This procedure is repeated iteratively until convergence. 

With the fixed $\lambda, \mu_1, \ldots, \mu_{l-1}, \mu_{l+1}, \ldots, \mu_L$, we define $\boldsymbol{\Xi}_l^{(n)} = \sum_{p \in \mathcal{K}} | \xi_p |^2 \mathbf{h}_p^{(n)} (\mathbf{h}_p^{(n)})^\text{H} + \lambda \mathbf{I} - \frac{1}{N_\text{c}} \sum_{q=1, q \neq l}^L \mu_q \mathbf{a}(\phi_q) \mathbf{a}^\text{H}(\phi_q)$. 
If $\boldsymbol{\Xi}_l^{(n)} - \frac{1}{N_\text{c}} \mu_l \mathbf{a}(\phi_l) \mathbf{a}^\text{H}(\phi_l)$ is invertible and $P_\text{T}^{(n)}(\phi_l) \geq P_\text{T,req}$ with $\mathbf{v}_k^{(n)} (\lambda, \{ \mu_1, \ldots, \mu_{l-1}, 0, \mu_{l+1}, \ldots, \mu_L\})$ for $k \in \mathcal{K}$, we have $\hat{\mu}_l=0$. Otherwise, the constraint \eqref{subeq:opt_c} should be satisfied with equality, which leads to $|\mathbf{a}^\text{H}(\phi_l) \mathbf{v}_k^{(n)}| = |\xi_k \sqrt{1+\beta_k} \mathbf{a}^\text{H}(\phi_l) (\boldsymbol{\Xi}_l^{(n)} - \tfrac{1}{N_c}\,\mu_l \mathbf{a}(\phi_l) \mathbf{a}^H(\phi_l))^{-1} \mathbf{h}_k |$. By the Sherman-Morrison formula, the following equation holds
\begin{align}
\label{eq:sherman-morrison}
    \big(\boldsymbol{\Xi}_l^{(n)} & - \tfrac{1}{N_c}\,\mu_l \mathbf{a}(\phi_l) \mathbf{a}^H(\phi_l)\big)^{-1} \\
    &\hspace{-0.1cm} = \big(\boldsymbol{\Xi}_l^{(n)}\big)^{-1} + \frac{\tfrac{1}{N_c}\,\mu_l\,\big(\boldsymbol{\Xi}_l^{(n)}\big)^{-1}\,\mathbf{a}(\phi_l)\,\mathbf{a}^H(\phi_l)\,\big(\boldsymbol{\Xi}_l^{(n)}\big)^{-1}}{1 - \tfrac{1}{N_c}\,\mu_l\,\mathbf{a}^H(\phi_l)\,\big(\boldsymbol{\Xi}_l^{(n)}\big)^{-1}\,\mathbf{a}(\phi_l)}. \nonumber
\end{align}
Inserting \eqref{eq:sherman-morrison} into \eqref{subeq:opt_c} and simplifying yield
\begin{align}
\label{eq:mu_eq}
    \sum_{k \in \mathcal{K}} (1+\beta_k) | \xi_k &\mathbf{a}^\text{H}(\phi_l) (\boldsymbol{\Xi}_l^{(n)})^{-1} \mathbf{h}_k |^2 \nonumber \\ 
    &=  \Big|1- \frac{1}{N_\text{c}} \mu_l \mathbf{a}^\text{H}(\phi_l) (\boldsymbol{\Xi}_l^{(n)})^{-1} \mathbf{a}(\phi_l) \Big|^2.
\end{align}
Then, the solution of \eqref{eq:mu_eq} is given by
\begin{align}
\label{eq:mu_sol}
    \mu_l = N_\text{c} \frac{\sqrt{P_{\text{T}, \text{req}}^{(n)}} \pm \sqrt{\sum\limits_{k \in \mathcal{K}} (1+\beta_k) | \xi_k \mathbf{a}^\text{H}(\phi_l) (\boldsymbol{\Xi}_l^{(n)})^{-1} \mathbf{h}_k |^2}}{\sqrt{P_{\text{T}, \text{req}}^{(n)}} \mathbf{a}^\text{H}(\phi_l) (\boldsymbol{\Xi}_l^{(n)})^{-1} \mathbf{a}(\phi_l)}. 
\end{align}
Since the Lagrange multiplier $\mu_l$ must satisfy the dual feasibility condition $\mu_l \geq 0$, the optimal $\mu_l$ is the smallest non-negative value in \eqref{eq:mu_sol}.

\subsection{Receive Beamformer Design}
\label{subsec:rx_beamformer_design}
To amplify the signals that are directly reflected at the focal angles $\{\phi_l\}_{l=1}^L$ configured during transmission, the BS employs a receive beamformer. If there are sensing targets on the focal angle $\phi_l$, from the definition of the received signal, the power of the received signal can be represented as
\begin{align}
    P_\text{R}(\phi_l) = \frac{|\bar{a}_l|^2 N_\text{sym}}{N_\text{c}} \sum_p \mathbf{u}^H \bar{\mathbf{A}}_l (\sum\limits_{k \in \mathcal{K}} \mathbf{v}_k^{(p)} (\mathbf{v}_k^{(p)})^\text{H}) \bar{\mathbf{A}}_l^\text{H} \mathbf{u},
\end{align}
% \begin{align}
%     P_\text{R}(\phi_l) &= \mathbb{E} (\| \bar{a}_l \mathbf{D}_{\text{I}}\left(\bar{f}_{\text{D},l}\right) \mathbf{F}_{N_\text{c}}^{-1} \mathbf{D}_{\text{R}}^*\left(\bar{\tau}_{l}\right) \mathbf{S}_l \mathbf{D}_{\text{v}}\left(\bar{f}_{\text{D},l}\right) \|_\text{F}^2) \nonumber \\
%     &= \mathbb{E}(\frac{|\bar{a}_l|^2}{N_\text{c}} \text{tr}(\mathbf{S}_l \mathbf{S}_l^\text{H})) \nonumber \\
%     &= \frac{|\bar{a}_l|^2 N_\text{sym}}{N_\text{c}} \sum_p \mathbf{u}^H \bar{\mathbf{A}}_l (\sum\limits_{k \in \mathcal{K}} \mathbf{v}_k^{(p)} (\mathbf{v}_k^{(p)})^\text{H}) \bar{\mathbf{A}}_l^\text{H} \mathbf{u},
% \end{align}
where $\bar{\mathbf{A}}_l = \mathbf{a}(\phi_l) \mathbf{a}^{\text{H}}(\phi_l)$.
Given the transmit beamformer $\{ \mathbf{v}_k^{(n)} \}$, after omitting the constant values, the receive beamformer design problem can be represented by
\begin{align}
\label{eq:opt_rx}
    \hspace{-0.25cm} 
    \max_{ \mathbf{u}} & & \frac{ \mathbf{u}^\text{H} \sum_{l=1}^{L} \big(|\bar{a}_l|^2 \sum_p \bar{\mathbf{A}}_l (\sum\limits_{k \in \mathcal{K}} \mathbf{v}_k^{(p)} (\mathbf{v}_k^{(p)})^\text{H}) \bar{\mathbf{A}}_l^\text{H} \big) \mathbf{u}}{\mathbf{u}^\text{H} \mathbf{u}}.
\end{align}
The objective in \eqref{eq:opt_rx} is a Rayleigh quotient, which is maximized when $\mathbf{u}$ is the eigenvector corresponding to the largest eigenvalue (i.e., the principal eigenvector) of $\mathbf{B}$, where $\mathbf{B} = |\bar{a}_l|^2 N_\text{sym} \sum_p \bar{\mathbf{A}}_l \Big(\sum\limits_{k \in \mathcal{K}} \mathbf{v}_k^{(p)} \big(\mathbf{v}_k^{(p)}\big)^\text{H}\Big) \bar{\mathbf{A}}_l^\text{H}$. Thus, the solution to \eqref{eq:opt_rx} is the principal eigenvector of $\mathbf{B}$, obtainable via standard eigenvalue decomposition.

\section{ICI Mitigation and Target Detection via DCFNet}
\label{sec:proposed_scheme}
In this section, we propose a deep learning-based ICI mitigation network, named DCFNet, to enhance target detection performance. Directly using sampled received signals as inputs is insufficient for extracting target features. To enable the model to extract ICI-free latent representation effectively, we employ a bank of DCFs. The proposed DCFNet then produces a RV map with significantly reduced ICI.

\subsection{DCFNet Architecture}
\begin{figure*}[t]
    \centering
    \includegraphics[draft=false, width= \if 1\mycmd 0.7 \else 1 \fi \linewidth]{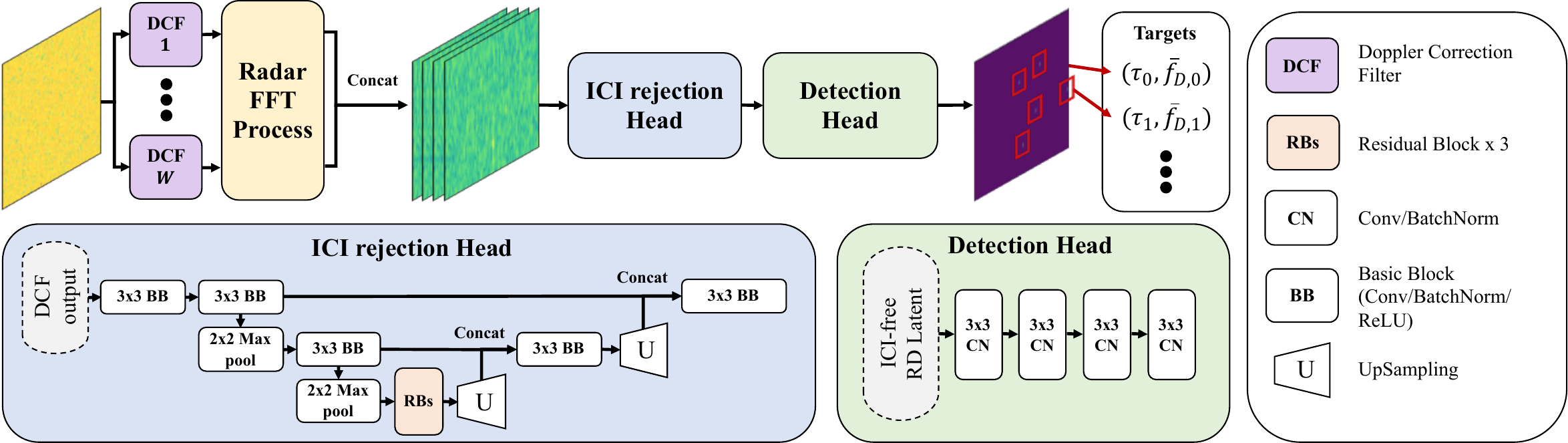}
    \caption{Architecture of DCFNet, which consists of three main components: (i) a bank of DCFs that shift ICI-robust regions to uncover targets buried by ICI, (ii) an ICI rejection head that suppresses ICI-related components and reconstructs a clean radar image using a U-shaped neural network, and (iii) a detection head that generates a confidence map indicating the likelihood of target presence in each range-velocity cell.}

    \label{fig:DCFNet_architecture}
    \vspace{-0.1cm}
\end{figure*}

Fig. \ref{fig:DCFNet_architecture} illustrates the architecture of DCFNet. DCFNet contains three key components: a bank of DCF, an ICI rejection head, and a detection head.
The received signal is sequentially processed through each component, ultimately representing the ranges and velocities of the targets on the RV map.

\noindent\textbf{Doppler correction filters:}
From the ICI term $\bar{\mathbf{Y}}_{\text{ICI},i}$, it can be observed that for a low-velocity target the ICI term can be negligible, i.e., $(\mathbf{D}_{\text{I}}\left(\bar{f}_{\text{D},i}\right)-\mathbf{I}_{N_\text{c}}) \approx \mathbf{0}$ and $\bar{\mathbf{Y}}_{\text{ICI},i} \approx \mathbf{0}$ while the ICI term caused by the high-mobility targets severely contaminates the RV map. 
Generally, in areas with relatively slow-moving targets, an ICI-robust region is formed, whereas in areas with relatively fast-moving targets, ICI-vulnerable regions are established.
We invert this phenomenon by putting aside the source of interference from fast-moving targets to slow-moving targets. Then, in contrast to the original scenario, fast-moving targets obscured by ICIs can be discovered, which ultimately increases the detection probability of targets in the high-velocity areas. This approach can be simply achieved by multiplying a DCF, i.e., $\mathbf{W}_w = \mathbf{D}_{\text{I}}^{*}\left(\bar{f}_{\text{DCF},w}\right), w \in \mathcal{W}$, where $\mathcal{W}$ is the filter index set. The filtered received signal can be represented by
\begin{align}\label{eq:filtered_received_signal_mat_form_divide}
    \mathbf{W}_w \mathbf{Y} = &\sum_{i \in \mathcal{P}} \bar{a}_i \mathbf{F}_{N_\text{c}}^{-1} \mathbf{D}_{\text{R}}^*\left(\bar{\tau}_{i}\right) \mathbf{S}_i \mathbf{D}_{\text{v}}\left(\bar{f}_{\text{D},i}\right) \hspace{1.3cm} \text{(continued)} \nonumber \\ &+ 
    \sum_{i \in \mathcal{P}} \bar{a}_i (\mathbf{D}_{\text{I}}\left(\bar{f}_{\text{D},i} - \bar{f}_{\text{DCF},w} \right)-\mathbf{I}_{N_\text{c}}) \mathbf{F}_{N_\text{c}}^{-1} \mathbf{D}_{\text{R}}^*\left(\bar{\tau}_{i}\right) \mathbf{S}_i \nonumber \\[-0.3cm] & \hspace{3.6cm} \cdot \mathbf{D}_{\text{v}}\left(\bar{f}_{\text{D},i}\right)
    + \mathbf{Z}_{\text{DCF}},
\end{align}
where $\mathbf{Z}_{\text{DCF}} = \mathbf{W}_w \mathbf{Z}$. In our approach, by modulating the value of $\bar{f}_{\text{DCF},w}$, we achieve the capability to methodically shift the focus of the ICI to a specific desired velocity interval.

\begin{figure*}[t]
\centering
\subfigure[Conventional]{
\includegraphics[draft=false, width= \if 1\mycmd 0.4 \else 0.5 \fi \columnwidth]{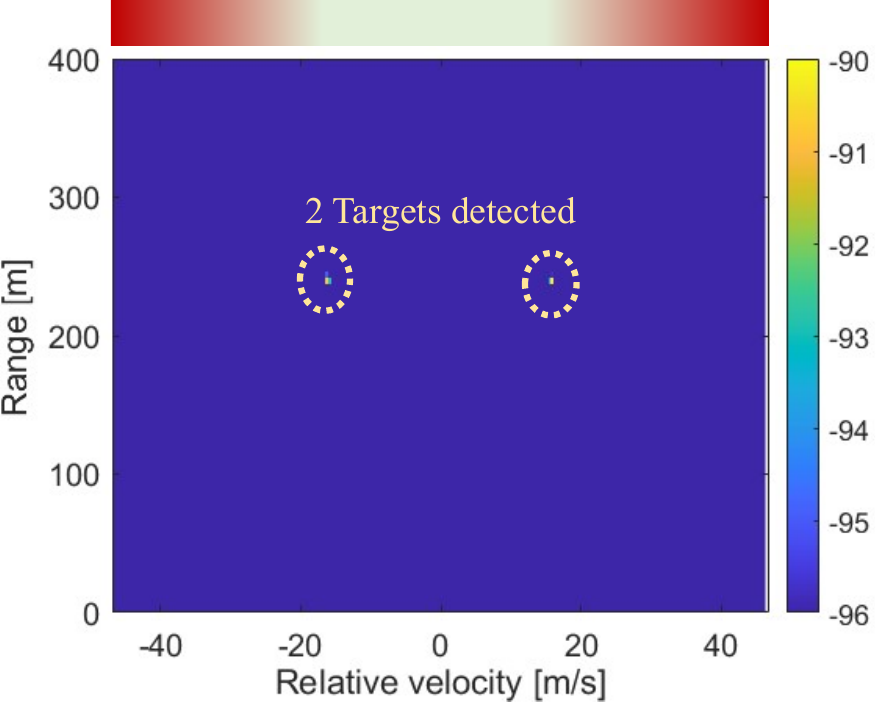}
}
\subfigure[$\bar{f}_{\text{DCF,0}} = 0.5$]{
\includegraphics[draft=false, width=\if 1\mycmd 0.4 \else 0.5 \fi\columnwidth]{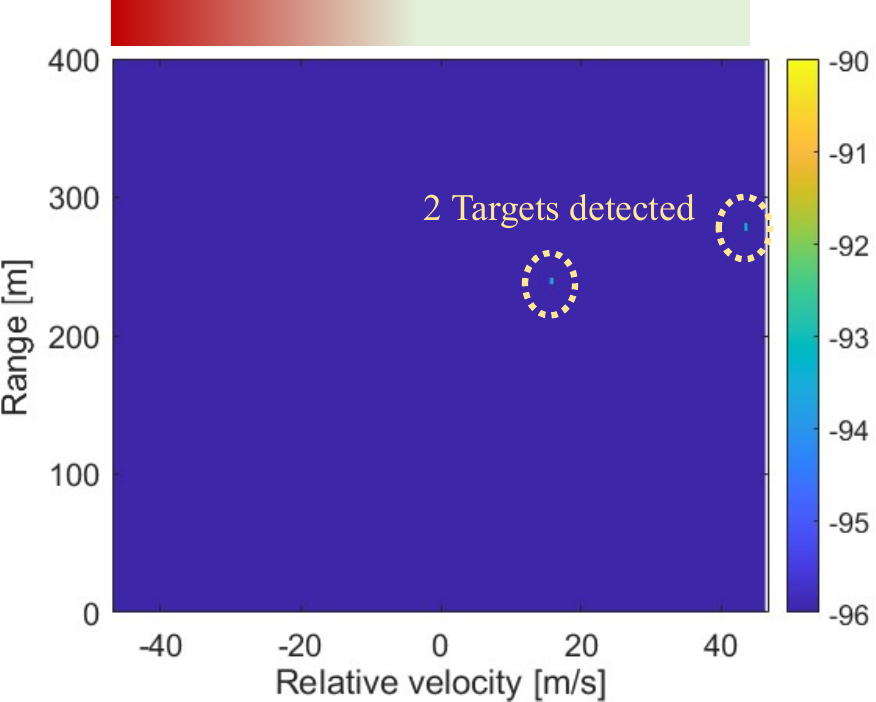}
}
\subfigure[$\bar{f}_{\text{DCF,1}} = -0.5$]{
\includegraphics[draft=false, width= \if 1\mycmd 0.4 \else 0.5 \fi \columnwidth]{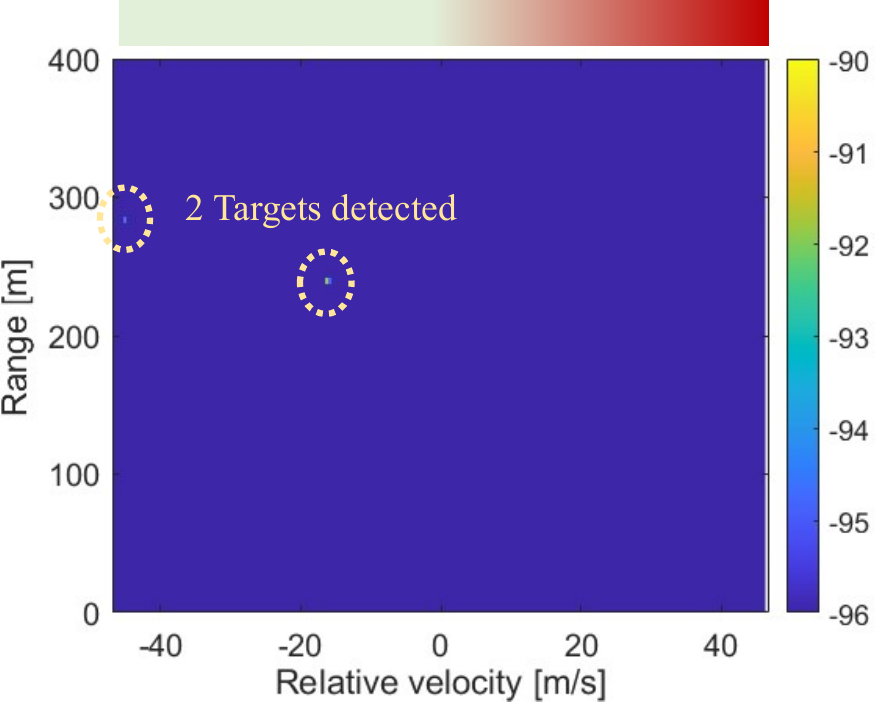}
}
\subfigure{
\raisebox{13pt}[\height][\depth]{\includegraphics[draft=false, width= \if 1\mycmd 0.4 \else 0.3 \fi \columnwidth]{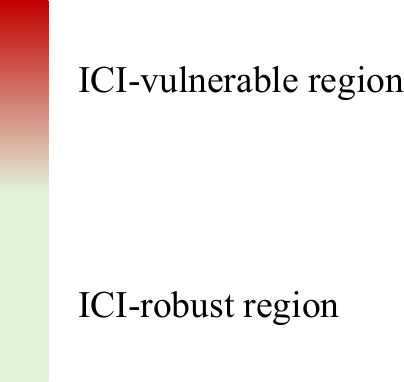}
}}
\vspace{-6pt}
\caption{The results of applying DCFs and radar FFT process sequentially. (a) Conventional 2D FFT process incurs severe ICI on the high-velocity areas. (b), (c) We apply two DCFs with $\bar{f}_{\text{DCF}, 0} = 0.5$ and $\bar{f}_{\text{DCF}, 1} = -0.5$, respectively. DCFs relocate the ICI-robust region to the high-velocity areas, revealing targets obscured by the ICIs.}
\label{fig:DCF_FFT}
\vspace{-6pt}
\end{figure*}

Figure \ref{fig:DCF_FFT} illustrates an example where four targets are located with different velocities and ranges. Directly applying the conventional 2D FFT, i.e., $\Pi(\mathbf{Y})$, cannot excavate the signals reflected by the fast-moving targets obscured by ICIs. We apply two DCFs with $\bar{f}_{\text{DCF}, 0} = 0.5$ and $\bar{f}_{\text{DCF}, 1} = -0.5$ and perform the radar FFT process, represented by $\Pi(\mathbf{W}_0 \mathbf{Y})$ and $\Pi(\mathbf{W}_1 \mathbf{Y})$, respectively. DCFs shift the ICI-robust regions from the low- to the high-velocity areas, revealing signals from fast-moving targets previously masked by ICI. Consequently, the low-velocity region becomes ICI-vulnerable, causing some previously detectable targets to be missed.

\noindent\textbf{ICI rejection head}: 
ICI rejection head aims to generate ICI-free radar RD latent from the filtered RV maps which passed through the DCFs and radar FFT process. Note that the DCFs shift the ICI source from high-velocity to low-velocity regions, thereby exposing signals masked by ICI, rather than directly mitigating ICI. Each filtered RV map serves as evidence to extract target features across the full velocity interval, while the ICI rejection head suppresses ICI-related components. To this end, we utilize a U-shaped architecture, commonly used in image reconstruction and generation tasks such as autoencoders and generative models \cite{Ronneberger15}. The result of ICI rejection head is given by
$\mathbf{Y}_\text{IRH} = \mathrm{IRH}(\mathrm{concat}(\Pi(\mathbf{W}_0 \mathbf{Y}), \Pi(\mathbf{W}_1 \mathbf{Y}), \ldots, \Pi(\mathbf{W}_{|\mathcal{W}|-1} \mathbf{Y})); \theta_{\text{IRH}}) $ $ \in \mathbb{R}^{d_{\text{IRH}} \times N_\text{c} \times N_\text{sym}},$
where $\text{IRH}(\cdot)$ denotes the ICI rejection head, which is composed of multiple basic blocks, max pooling, upsampling, and residual blocks. The operation $concat(\cdot)$ denotes channel-wise concatenation.  $d_\text{IRH}$ is the number of output feature channels, and $\theta_{\text{IRH}}$ represents the learnable parameters of the ICI rejection head. Each basic block consists of a $3\times3$ convolutional layer, batch normalization, and ReLU activation.

The ICI rejection head first extracts feature maps from the filtered RV maps and downsamples them through three basic blocks and two max-pooling layers. The max-pooling emphasizes key features, such as target information, while suppressing ICI-related components, thereby enhancing target discrimination. Residual blocks are integrated to mitigate overfitting and enable deeper architectures without performance degradation.
A key challenge, however, is the resolution loss, which is critical for precise target localization. The two max-pooling layers reduce the feature map size to $N_\text{c}/4 \times N_\text{sym}/4$, causing a fourfold loss in both range and velocity resolution. To address this, the feature maps are upsampled and concatenated with earlier features twice, restoring fine-grained details at the original resolution in the final RV map. 

\vspace{0.18cm}
\noindent\textbf{Detection head}: 
Our detection head is inspired by the object detection framework in the AI-based object detection and segmentation work \cite{Rebut22}. The detection head takes ICI-free RD latent as an input and passes it through four sets of $3\times3$ convolutional neural net and batch normalization. Finally, the feature map is processed with a sigmoid activation, yielding a confidence map that indicates the likelihood of a target's presence at each pixel. The overall process can be denoted as $\mathbf{P}_{\text{DCF}} = \text{DH}(\mathbf{Y}_{\text{IRH}}; \theta_{\text{DH}}) \in [0,1]^{N_\text{c} \times N_\text{sym}}$, where $\text{DH}(\cdot)$ is the detection head, and $\theta_{\text{DH}}$ is the learnable parameters.

\subsection{Training Process}

We train the ICI rejection head and detection head using supervised learning. Thus, we first define a dataset $\mathcal{D}$ with the size of $D$. In the dataset, each data point consists of the output of DCF and radar FFT process, i.e., $\{ \Pi(\mathbf{W}_w \mathbf{Y}^{(j)}) \}_{w=0}^{|\mathcal{W}|-1}\}$, and a ground truth detection map $\hat{\mathbf{P}}_{\text{G.T.}}^{(j)}$ where $j$ indicates the data index in the dataset $\mathcal{D}$.
% In order to generate a ground truth detection map, we render $\hat{\mathbf{P}}_{\text{G.T.}}^{(j)}$ to mimic the ICI-free RV map. This matrix can be constructed based on the first term in \eqref{eq:radar_image}. However, it includes both the target locations and the sidelobe effects with the latter clearly outside the scope of our interest. 
In order to generate a ground truth detection map, we create a binary matrix $\hat{\mathbf{P}}_{\text{G.T.}}^{(j)} \in \{0,1\}^{N_\text{c} \times N_\text{sym}}$ wherein only the indices corresponding to the presence of targets are assigned a value of one, with all other indices set to zero. This definition can be denoted as 
\begin{align}
\label{eq:ground_truth}
    \hspace{-0.3cm}
    [\hat{\mathbf{P}}_{\text{G.T.}}^{(j)}]_{n,\mu} = \begin{cases}
        1, ~ \text{if} ~ (n, \mu) = (\ceil{\bar{\tau}_i^{(j)} N_\text{c}}, \ceil{\bar{f}_{\text{D},i}^{(j)} N_\text{sym}}), \forall i, \\
        0, ~ \text{otherwise},
    \end{cases}
    \hspace{-0.5cm}
\end{align}
where $\ceil{\cdot}$ denotes ceil function that returns the smallest integer value which is greater than or equal to a number. 

It is worth noting that target distribution is generally very sparse, and thus there exists a significant data imbalance in target classification. To handle the problem, we apply a focal loss \cite{Lin17} to all pixels for the classification. When DCFNet predicts a confidence map $\mathbf{P}_\text{DCF}^{(j)}$, its training loss reads:
\begin{align}\label{eq:focal_loss}
    \hspace{-0.3cm}
    \mathcal{L}_\text{focal}(\mathbf{P}_\text{DCF}^{(j)}, \hat{\mathbf{P}}_{\text{G.T.}}^{(j)})= - \!\!\!\sum_{\mu=0}^{N_\text{sym}-1} \!\sum_{n=0}^{N_\text{c}-1} (1-p_{n,\mu}^{(j)})^\gamma \log(p_{n,\mu}^{(j)}),
\end{align}
where $\gamma$ is the tunable focusing parameter, and $p_{n,\mu}^{(j)}$ equals $[\mathbf{P}_{\text{DCF}}^{(j)}]_{(n,\mu)}$ if $[\hat{\mathbf{P}}_{\text{G.T.}}^{(j)}]_{(n,\mu)} = 1$, and $1 - [\mathbf{P}_{\text{DCF}}^{(j)}]_{(n,\mu)}$ otherwise.

Our aim is to update the network parameters $\theta_{\text{IRH}}$ and $\theta_{\text{DH}}$. In each training epoch, we randomly sample a mini-batch $\mathcal{M}$ from the dataset $\mathcal{D}$ and update the network parameters by the stochastic gradient descent method. The detailed training algorithm for DCFNet is presented in Algorithm \ref{alg:train}. 

\begin{algorithm}[t]
\algnewcommand{\Initialize}[1]{%
  \State \textbf{Initialize:}
  \Statex \hspace*{\algorithmicindent}\parbox[t]{.8\linewidth}{\raggedright #1}
}
\caption{Training algorithm for DCFNet}
\label{alg:train}
\begin{algorithmic}[1]
\State \textbf{Input:} Dataset $\mathcal{D}$ consisting of $\{ \Pi(\mathbf{W}_w \mathbf{Y}^{(j)}) \}_{w=0}^{|\mathcal{W}|-1}$ and $ \{ \bar{f}_{\text{D},i}^{(j)}, \bar{\tau}_{i}^{(j)} \}_{i\in \mathcal{P}, j=0,\ldots, D-1}$, maximum number of epochs $T$.
\State \textbf{Output:} Optimized network parameters $\theta_{\text{IRH}}$, $\theta_{\text{DH}}$.
    \vspace{3pt}

\State Initialize $\theta_{\text{IRH}}$, $\theta_{\text{DH}}$.
\For{$t=1$ to $T$}
\State Sample mini-batch $\mathcal{M}$ from the dataset $\mathcal{D}$.
\State Construct $\hat{\mathbf{P}}_{\text{G.T}}^{(m)}$ from \eqref{eq:ground_truth} for all $m \in \mathcal{M}$.
\State Compute $\mathcal{L}_\text{focal}(\mathbf{P}_\text{DCF}^{(m)}, \hat{\mathbf{P}}_{\text{G.T.}}^{(m)})$ from \eqref{eq:focal_loss} for all $m \in \mathcal{M}$.
\State \multiline{Update the network parameters $\theta_{\text{IRH}}$, $\theta_{\text{DH}}$ using $\mathcal{L}_\text{focal}(\mathbf{P}_\text{DCF}^{(m)}, \hat{\mathbf{P}}_{\text{G.T.}}^{(m)})$.}
\EndFor
\end{algorithmic}
\end{algorithm}
\vspace{-0.2cm}

% \begin{algorithm}[t]
% \algnewcommand{\Initialize}[1]{%
%   \State \textbf{Initialize:}
%   \Statex \hspace*{\algorithmicindent}\parbox[t]{.8\linewidth}{\raggedright #1}
% }
% \caption{ML-based iterative algorithm}
% \label{alg:ML_detection}
% \begin{algorithmic}[1]
% \Initialize{
%     $\mathbf{y}_\text{vec}^{(0)}$, $\eta$;
%     }
%     \vspace{3pt}
% \Repeat 
%     \State Compute $\bar{a}_p^\text{opt}, \bar{f}_{\text{D},p}^{\text{opt}}, \bar{\tau}_p^{\text{opt}}$ in \eqref{eq:a_opt} and \eqref{eq:GLRT_argmax};
%     \State Compute $\mathcal{L}(\mathbf{y}_\text{vec}^{(p)})$ in \eqref{eq:GLRT_log};
%     \If{$\mathcal{L}(\mathbf{y}_\text{vec}^{(p)}) > \eta$}
%         \State $\mathbf{y}_\text{vec}^{(p+1)} = \mathbf{y}_\text{vec}^{(p)} - \bar{a}_p \Xi (\bar{f}_{\text{D}, p}, \bar{\tau}_p)$;
%     \Else
%         \State Break;
%     \EndIf
% \Until{$p < p_\text{max}$}
% \end{algorithmic}
% \end{algorithm}

\subsection{ML-based Local Refinement for Multi-target ranges and velocities Estimation}
While the ranges and velocities of targets can be estimated from $\mathbf{P}_\text{DCF}$, the resolution of both range and velocity is inherently limited by the constraints of the 2D FFT process.
%In this subsection, we further improve the accuracy for range and velocity estimations beyond the curse of resolution in 2D FFT methods by developing a local refinement and seamlessly integrating it with DCFNet. 
In this subsection, we propose DCFNet-LR which utilizes GLRT to improve the accuracy for range and velocity estimations beyond the curse of resolution in 2D FFT methods.

The target ranges and velocities in ML-based approaches are determined by the following optimization problem:
\begin{align}
    &\bar{f}_\text{D}^\text{opt}, \bar{\tau}^\text{opt} \\ &=  \argmin_{\bar{f}_\text{D}, \bar{\tau}} \Big\| \mathbf{Y} - \sum_{i \in \mathcal{P}} \bar{a}_i \mathbf{D}_{\text{I}}\left(\bar{f}_{\text{D},i}\right) \mathbf{F}_{N_\text{c}}^{-1} \mathbf{D}_{\text{R}}^*\left(\bar{\tau}_{i}\right) \mathbf{S}_i \mathbf{D}_{\text{v}} \hspace{-2pt}  \left(\bar{f}_{\text{D},i}\right) \Big\|_\text{F}^2. \nonumber
\end{align}
However, the number of targets is evidently unknown variables in the receiver side. In addition, it is intractable to find the global optimal solution due to the non-convexity characteristic of the objective function. To this end, We first detect sensing targets coarsely via DCFNet and estimate the much accurate ranges and velocities of targets with GLRT.

For ease of notation, we first vectorize the matrix representation of the receive signal as follows:
\begin{align}
    & \mathbf{y}_\text{vec} = \left[ \left[ \left(\mathbf{Y}\right)^\text{T} \right]_{(:,0)}, \ldots, \left[ \left(\mathbf{Y}\right)^\text{T} \right]_{(:,N_\text{sym}-1)} \right]^\text{T}, \\
    & \Psi_{\mu}(\bar{f}_{\text{D},p}, \bar{\tau}_p)= \mathbf{D}_{\text{I}}\left(\bar{f}_{\text{D},p}\right) \mathbf{F}_{N_\text{c}}^{-1} \mathbf{D}_{\text{R}}^*\left(\bar{\tau}_{p}\right) \left[ \mathbf{S}_{p} \right]_{(:,\mu)} e^{j2\pi \bar{f}_{\text{D},p} \alpha \mu}, \\
    & \Xi(\bar{f}_{\text{D},p}, \bar{\tau}_p) = [\Psi_{0}^{\text{T}}(\bar{f}_{\text{D},p}, \bar{\tau}_p), \ldots, \Psi_{N_\text{sym}-1}^{\text{T}}(\bar{f}_{\text{D},p}, \bar{\tau}_p)]^\text{T}.
\end{align}
Then, GLRT can be written by
\begin{align}
    \mathcal{L}_\text{GLRT}(\mathbf{y}_\text{vec}) = \frac{\max\limits_{\bar{a}_p, \bar{f}_{\text{D},p}, \bar{\tau}_p} p(\mathbf{y}_\text{vec} | \mathcal{H}_1;\bar{a}_p, \bar{f}_{\text{D},p}, \bar{\tau}_p )}  {p(\mathbf{y}_\text{vec} | \mathcal{H}_0)}  \underset{{\mathcal{H}_0}}{\overset{\mathcal{H}_{1}}{\gtrless}} \tilde{\eta},
\end{align}
for threshold $\tilde{\eta}$. Since we assume the complex additive Gaussian noise, the GLRT can be represented by
\begin{align}
\label{eq:GLRT_gauss}
    &\mathcal{L}_\text{GLRT}(\mathbf{y}_\text{vec})  \\ 
    &= \frac{\exp \big(- \min\limits_{\bar{a}_p, \bar{f}_{\text{D},p}, \bar{\tau}_p} \left\| \mathbf{y}_\text{vec} - \Bar{a}_p \Xi(\bar{f}_{\text{D},p}, \bar{\tau}_p) \right\|_2^2 / \sigma^2 \big)}  {\exp \left( -N_\text{c}N_\text{sym}/\sigma^2 \right)}  \underset{{\mathcal{H}_0}}{\overset{\mathcal{H}_{1}}{\gtrless}} \tilde{\eta}. \nonumber
\end{align}
Then, we take the log of \eqref{eq:GLRT_gauss}, which can be represented by
\begin{align}
\label{eq:GLRT_log}
    &\mathcal{L}_\text{GLRT, log}(\mathbf{y}_\text{vec}) =   \\ 
    & \hspace{0.3cm}  N_\text{c}N_\text{sym}/\sigma^2 - \min_{\bar{a}_p, \bar{f}_{\text{D},p}, \bar{\tau}_p} \left\| \mathbf{y}_\text{vec} - \Bar{a}_p \Xi(\bar{f}_{\text{D},p}, \bar{\tau}_p) \right\|_2^2 / \sigma^2  \underset{{\mathcal{H}_0}}{\overset{\mathcal{H}_{1}}{\gtrless}} \eta, \nonumber
\end{align}
where $\eta = \log \tilde{\eta}$.
To reduce the optimization parameter and carry out the minimization, we fix the variables $\bar{f}_{\text{D},p}, \bar{\tau}_p$ and minimize \eqref{eq:GLRT_log} with respect to $\bar{a}_p$, which yields 
\begin{align}
\label{eq:a_opt}
    \bar{a}_p^\text{opt} = (\mathbf{y}_\text{vec})^\text{T} \Xi^*(\bar{f}_{\text{D},p}, \bar{\tau}_p) / \Xi^\text{T}(\bar{f}_{\text{D},p}, \bar{\tau}_p) \Xi^*(\bar{f}_{\text{D},p}, \bar{\tau}_p).
\end{align}
Substituting $\bar{a}_p^\text{opt}$ into the minimization in \eqref{eq:GLRT_log}, we can obtain the optimal solution as follows:
\begin{align}
\label{eq:GLRT_argmax}
    \hspace{-0.35cm}
    \bar{f}_{\text{D},p}^{\text{opt}}, \bar{\tau}_p^{\text{opt}} = \argmax_{\bar{f}_{\text{D},p}, \bar{\tau}_p} ~(\mathbf{y}_\text{vec})^\text{H} \Xi (\bar{f}_{\text{D},p}, \bar{\tau}_p)\Xi^\text{H}(\bar{f}_{\text{D},p}, \bar{\tau}_p) \mathbf{y}_\text{vec}.
    \hspace{-0.3cm}
\end{align}
Due to its non-concave nature, the objective function in \eqref{eq:GLRT_argmax} poses challenges in reliably finding the global maximum.

\begin{table}
    \centering
    \caption{OFDM System Parameters}
    \label{table:spec_AI_radcom}
    \resizebox{\if 1\mycmd 0.6 \else 0.95 \fi \columnwidth}{!}{
    \begin{tabular}{ccc}
    \toprule
         \textbf{Parameter} & \textbf{Symbol} & \textbf{Value} \\ \midrule[\heavyrulewidth]\midrule[\heavyrulewidth]
         \arrayrulecolor{lightgray}
         Number of transmit antennas & $N_\text{T}$ & $20$ \\ \cline{1-3}
         Number of receive antennas & $N_\text{R}$ & $20$ \\ \cline{1-3}
         Transmit power & $P_\text{BS}$ & $30~\text{dB}$ \\ \cline{1-3}
         Carrier frequency & $f_\text{c}$ & $60~\text{GHz}$ \\ \cline{1-3}
         Bandwidth & $B$ & $50~\text{MHz}$ \\ \cline{1-3}
         Number of subcarriers & $N_\text{c}$ & $2048$ \\ \cline{1-3}
         Subcarrier spacing & $\Delta f = B / N_\text{c}$ & $24.41~\text{kHz}$ \\ \cline{1-3}
         OFDM signal duration & $T = 1/ \Delta f$ & $40.96~ \mu\text{s}$ \\ \cline{1-3}
         Number of OFDM signals & $N_\text{sym}$ & $64$ \\ \cline{1-3}
         Total OFDM signal duration & $T_\text{OFDM} = \alpha T$ & $51.2~\mu \text{s}$ \\ \cline{1-3}
         Unambiguous range & $R_\text{max}$ & $6144~\text{m}$ \\ \cline{1-3}
         Unambiguous velocity & $v_\text{rel,max}$ & $\pm 24.41~\text{m/s}$ \\ \cline{1-3}
         Range resolution (FFT) & $\Delta R$ & $3~\text{m}$ \\ \cline{1-3}
         Velocity resolution (FFT) & $\Delta v_\text{rel}$ & $0.76~\text{m/s}$ \\ \cline{1-3}
         Number of sub-grid & $M_\text{c}, M_\text{sym}$ & 64 \\
         \arrayrulecolor{black}
    \bottomrule
    \end{tabular}
    }
\label{table:parameter_symbol_value_AI_radcom}
\vspace{-6pt}
\end{table}

ML-based methods address the optimization in \eqref{eq:GLRT_argmax} by exhaustively evaluating the objective function over the entire range-velocity space. To match the resolution of FFT-based methods, the ML detector performs a search over a coarse grid with $N_\text{c}$ range bins and $N_\text{sym}$ Doppler bins resulting in $N_\text{c}N_\text{sym}$ candidate cells. To further achieve super-resolution, each coarse bin is uniformly partitioned into $M_\mathrm{c}$ and $M_\text{sym}$ finer bins along the range and velocity dimensions, respectively. Then, the ML detector must evaluate the test statistic at $N_\text{c} M_\text{c} N_\text{sym} M_\text{sym}$ points, with each evaluation requiring approximately $\mathcal{O}(N_\text{c}^{2}N_\text{sym}^{2})$ floating-point operations. This results in a total per-iteration computational complexity of $\mathcal{O}(N_\text{c}^{3}M_\text{c} N_\text{sym}^{3}M_\text{sym})$.

% ML-based methods solve the problem in \eqref{eq:GLRT_argmax} by evaluating all the values of the objective function over the full range-velocity plane. To obtain the same resolution of FFT-based methods, ML-based methods search over the coarse FFT map which contains $N_\mathrm{c}$ range bins (one per sub-carrier) and $N_\mathrm{sym}$ Doppler bins (one per OFDM symbol), i.e., $N_\mathrm{c}N_\mathrm{sym}$ candidate cells. To further obtain the super-resolution, each coarse cell is uniformly subdivided into $M_\mathrm{c}$ and $M_\mathrm{sym}$ finer bins along the range and Doppler axes, respectively. The conventional ML detector must therefore evaluate the test statistic at $N_\text{c} M_\text{c}, N_\text{sym} M_\mathrm{text}$ points, and a single evaluation costs $\mathcal{O}\!\bigl(N_\mathrm{c}^{2}N_\mathrm{sym}^{2}\bigr)$ flops, so the per-iteration complexity scales as $\mathcal{O}\!\bigl(N_\mathrm{c}^{3}M_\mathrm{c}\,N_\mathrm{sym}^{3}M_\mathrm{sym}\bigr)$.

DCFNet-LR circumvents this challenge on huge computation burden. From the DCFNet output matrix $\mathbf{P}_\mathrm{DCF}$ we retain only the $p_{\max}$ cells whose confidence exceeds a threshold $\delta$, i.e.,
$\{(\bar{f}_{\mathrm{D},p},\bar{\tau}_{p})\mid[\mathbf{P}_\mathrm{DCF}]_{(\bar{f}_{\mathrm{D},p},\bar{\tau}_{p})}>\delta\}$.  
Each surviving cell is ranked by confidence and refined on a local $M_\mathrm{c}\!\times\!M_\mathrm{sym}$ sub-grid.  
Accordingly, the number of GLRT evaluations collapses from $N_\mathrm{c}N_\mathrm{sym}M_\mathrm{c}M_\mathrm{sym}$ to merely $M_\mathrm{c}M_\mathrm{sym}$.  
Moreover, because DCFNet already isolates individual targets, these refinements are performed once and in parallel, whereas a classical ML receiver must repeatedly re-estimate the grid after cancelling previously detected strong echoes.

% The whole algorithm is summarized in Algorithm \ref{alg:ML_detection}.

\section{Simulation Results} \label{sec:sim}
% In this section, we verify that DCFNet significantly improves radar target detection performance.

\subsection{Simulation Setups}
For simulations, the radar parameters are chosen from the literature \cite{Keskin21, Noh23}, and important system parameters are summarized in Table \ref{table:spec_AI_radcom}. 
It is worthy noting that the proposed method is not limited to the selected parameters. Four communication users are located at a range of $40~\text{m}$ and are moving at walking speed. We define the total transmit power budget and sensing power as $P_\text{BS} = \sum_{n=0}^{N_\text{c}-1} P_\text{BS}^{(n)}$ and $P_{\text{T}, \text{req}} = \sum_{n=0}^{N_\text{c}-1} P_{\text{T}, \text{req}}^{(n)}$, respectively. The channel from the BS to each user is modeled based on 3GPP TR 38.901 \cite{3gpp38901}. The noise variance $\sigma_k^2$ is set to $-97$ dBm for all the users and the BS. 
All simulations are implemented on an AMD Ryzen\textsuperscript{\texttrademark} 9 5800X processor and NVIDIA RTX 4090 GPU.

\begin{figure}[t]
    \centering
    \includegraphics[draft=false, width= \if 1\mycmd 0.7 \else 0.9 \fi \linewidth]{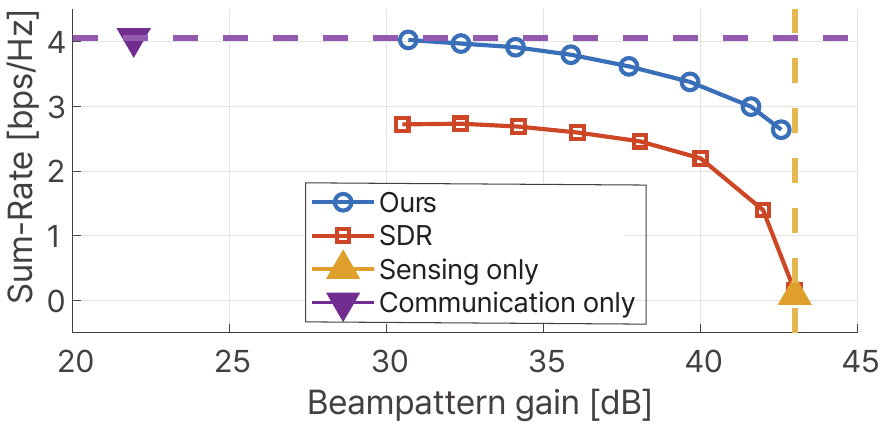}
    \vspace{-0.2cm}
    \caption{\mbox{Sum-rate per beampattern gain for DCFNet and the baseline methods.}}
    \label{fig:SR_BG}
\end{figure}

\begin{figure}[t]
    \centering
    \includegraphics[draft=false, width= \if 1\mycmd 0.7 \else 1 \fi \linewidth]{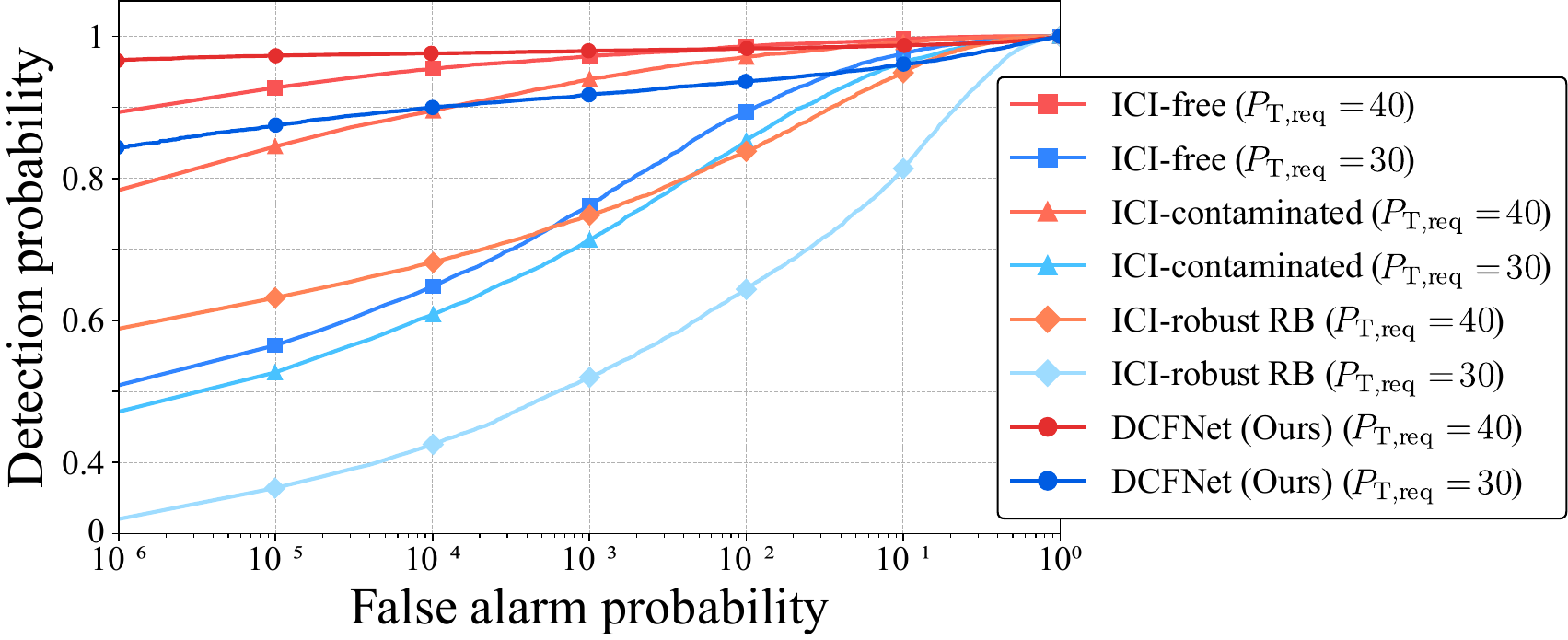}
    \vspace{-0.2cm}
    \caption{ROC curve for DCFNet and the baseline methods.}
    \label{fig:RoC}
\end{figure}

\begin{figure*}[t]
\centering
\subfigure[ICI-contaminated]{
\includegraphics[draft=false, width=0.35\columnwidth]{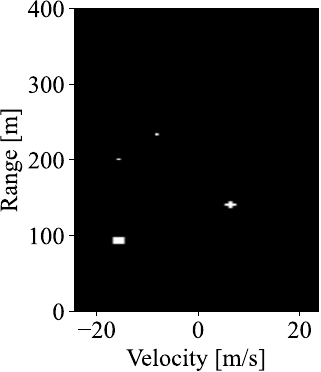}
}%
\subfigure[ICI-robust RB]{
\includegraphics[draft=false, width=0.35\columnwidth]{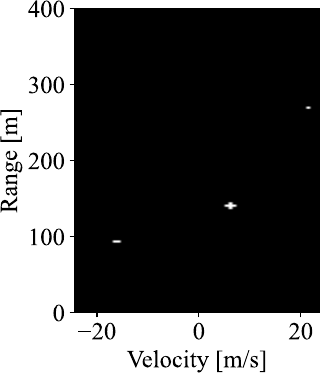}
}
\subfigure[ICI-free]{
\includegraphics[draft=false, width=0.35\columnwidth]{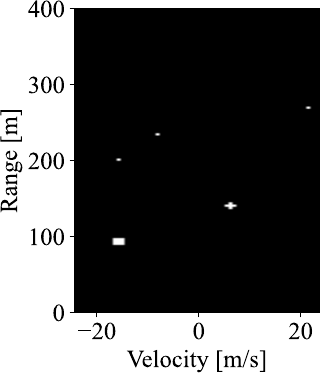}
}
\subfigure[DCFNet (Ours)]{
\includegraphics[draft=false, width=0.35\columnwidth]{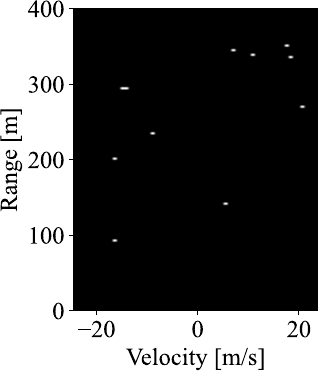}
}
\subfigure[Ground-truth]{
\includegraphics[draft=false, width=0.425\columnwidth]{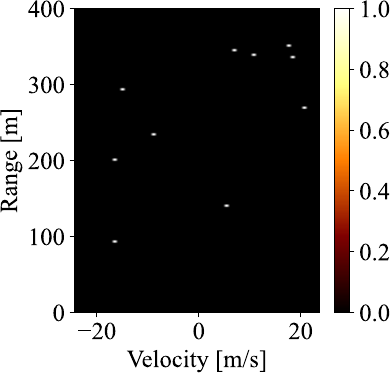}
}
\vspace{-6pt}
\caption{Comparison of detection maps under different interference and processing scenarios. We set false alarm probability of $3.0 \times 10^{-3}$ for both CFAR detector and DCFNet.}
\label{fig:RD_maps}
\vspace{-8pt}
\end{figure*}

\begin{figure*}[t]
\centering
\subfigure[DCFNet (Ours)]{
\includegraphics[draft=false, width=0.43\columnwidth]{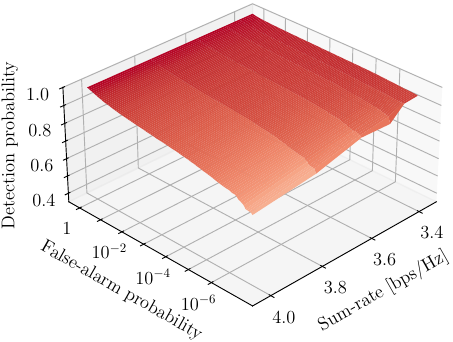}
}
\subfigure[ICI-free]{
\includegraphics[draft=false, width=0.43\columnwidth]{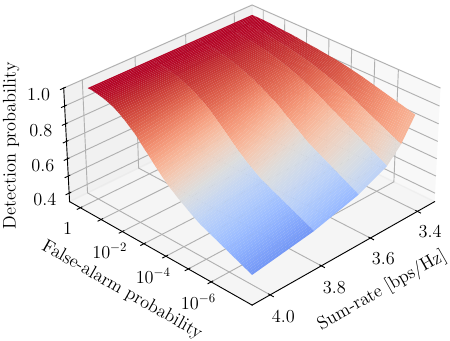}
}
\subfigure[ICI-contaminated]{
\includegraphics[draft=false, width=0.43\columnwidth]{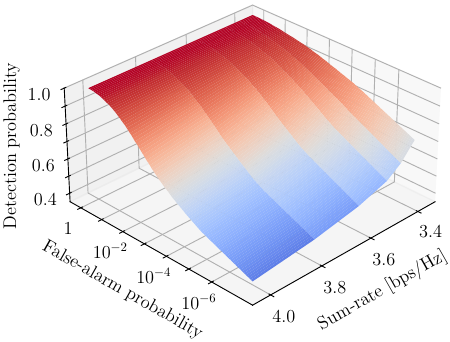}
}
\subfigure[ICI-robust RB]{
\includegraphics[draft=false, width=0.53\columnwidth]{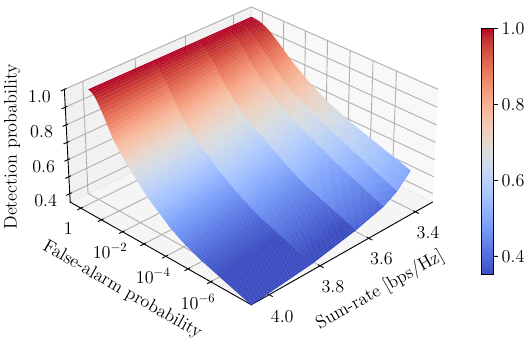}
}
\vspace{-6pt}
\caption{Detection probability with respect to the false-alarm probability and communication sum-rate.}
\label{fig:surf_sr_dp_fp}
\vspace{-8pt}
\end{figure*}

\subsection{Analysis on the Proposed Beamformers}
We consider three different transmit beamformer baselines.
\begin{itemize}[leftmargin=*]
    \item \textbf{SDR}: It adopts transmit covariance matrix and omits rank-1 constraint to relax the optimization problem into a convex form.
    \item \textbf{Sensing only}: It maximizes the beampattern gain without any communication constraint.
    \item \textbf{Communication only}: It solves the optimization problem in \eqref{eq:opt} without sensing constraint \eqref{subeq:opt_b}.
\end{itemize}

In Fig. \ref{fig:SR_BG}, the trade-off between beampattern gain and sum-rate is evaluated under varying beampattern gain constraint. Since the sensing-only and communication-only schemes are unconstrained in their respective domains, each results in a single (beampattern gain, sum-rate) pair. The result confirms that the proposed transmit beamformer consistently outperforms the SDR-based approach in both sensing and communication performance. This is mainly because the SDR method extracts a beamforming vector from the transmit covariance matrix via eigenvalue decomposition or Gaussian randomization, which typically results in a vector that falls short of fully utilizing the available power budget. Consequently, it ultimately leads to reduced ISAC performance.

\subsection{Detection Performance of DCFNet Compared to Conventional FFT-based Detection Methods}
We use the following FFT-based baseline methods:
\begin{itemize}[leftmargin=*]
    \item \textbf{ICI-contaminated}: It does not adopt any ICI rejection method. Thus, it utilizes $\bar{\mathbf{Y}}$ in \eqref{eq:radar_image} for target detection, which is clearly contaminated by ICI.
    \item \textbf{ICI-free}: The 2-D FFT method applied to the ICI-free received data, i.e., $\bar{\mathbf{Y}}$ in \eqref{eq:radar_image} with $\bar{\mathbf{Y}}_{\text{ICI},i} = 0, \forall i$. 
    \item \textbf{ICI-robust receive beamformer (RB)} \cite{Noh23}: It optimizes receive beamformer via beam synthesis while rejecting ICI.
\end{itemize}
Since prior works did not propose a concrete detection mechanism, we adopt a conventional constant false alarm rate (CFAR) detector for all baseline schemes. This allows a fair comparison in terms of detection probability and false alarm probability, which are defined as follows:
\begin{itemize}[leftmargin=*]
    \item \textbf{Detection probability ($P_\text{D}$)}: the ratio of correctly detected targets to the total number of actual targets in the scene.
    \item \textbf{False alarm probability ($P_\text{FA}$)}: the ratio of falsely detected targets (i.e., detections not corresponding to any real target) to the total number of sensing image cells.
\end{itemize}
We evaluate the metrics over various CFAR thresholds and plot receiver operating characteristic (ROC) curves to assess sensitivity-selectivity trade-offs.

In Fig.~\ref{fig:RoC}, we depict the ROC curves of all FFT-based schemes with the proposed transmit and receive beamformers obtained in Sec.~\ref{subsec:tx_beamformer_design} and Sec.~\ref{subsec:rx_beamformer_design}. DCFNet consistently outperforms all baselines across various false alarm probabilities, even under the stringent radar power constraint of $P_\text{T,req} = 40$~dB. Notably, it achieves $P_\text{D} \approx 0.97$ even when $P_\text{FA}$ drops below $10^{-5}$. Unlike ICI-contaminated scheme, which significantly degrades at low false alarm regimes, DCFNet shows a substantial improvement, demonstrating strong robustness against Doppler-induced ICI.

Interestingly, DCFNet outperforms even the ICI-free scheme in terms of detection probability, despite the latter being unaffected by ICI.
This is attributed to the well-trained detection head in DCFNet, which effectively distinguishes true targets from background noise. In contrast, the ICI-robust RB scheme, which suppresses ICI through receiver beamformer optimization, performs worse than the ICI-contaminated case. This counterintuitive result arises from the limitations of prior ICI rejection techniques \cite{Noh23, hakobyan18}, which primarily suppress ICI for targets with positive velocities. However, this asymmetry inadvertently degrades resolution in the negative velocity region, severely impacting overall detection performance.

Fig.~\ref{fig:RD_maps} provides qualitative insight into the performance of different processing schemes by visualizing their corresponding RV detection maps. All methods are evaluated under an identical transmit beamformer configuration with a beampattern gain constraint of $P_{\text{T}, \text{req}} = 30~\text{dB}$, and $P_\text{FA} = 3.0 \times 10^{-3}$.

In Fig.~\ref{fig:RD_maps}(a), the ICI-contaminated baseline misses all targets beyond $250~\text{m}$. Doppler-induced ICI raises the noise floor, driving the effective target SNR so low that distant echoes vanish.  
The ICI-robust RB in Fig.\ref{fig:RD_maps}(b) mitigates this effect only on the positive velocity region; it still fails to resolve targets in the negative velocity, revealing the limit of its asymmetric suppression.
In contrast, Fig.~\ref{fig:RD_maps}(c) shows that the ICI-free result produces clean and well-localized target responses by ignoring the ICI effects. Although this represents an idealized upper bound for FFT-based processing, the proposed DCFNet output in Fig.~\ref{fig:RD_maps}(d) surpasses even this bound: its data-driven detection head suppresses background clutter, eliminates false alarms, and removes main-lobe spreading, producing the most accurate and visually pristine detection map among all the methods.

Fig.~\ref{fig:surf_sr_dp_fp} illustrates the detection probability as a joint function of false-alarm probability and communication sum-rate for the FFT-based schemes. The proposed DCFNet in Fig.\ref{fig:surf_sr_dp_fp}(a) maintains high detection performance across the domain, achieving $P_\text{D} > 0.9$ even at low false-alarm levels ($P_\text{FA} < 10^{-5}$) while supporting sum-rates over $3.8~ \text{bps/Hz}$. The surface exhibits smooth and gradual degradation, indicating that DCFNet maintains robustness under both interference and communication throughput constraints.
In contrast, the ICI-free and ICI-contaminated schemes shown in Figs.~\ref{fig:surf_sr_dp_fp}(b) and (c) exhibit noticeable declines in detection probability as the sum-rate increases or the false-alarm constraint becomes stricter. This highlights the limit of CFAR detector, the vulnerability to ICI, and lack of adaptive mitigation mechanisms. The ICI-robust RB in Fig.~\ref{fig:surf_sr_dp_fp}(d) shows the poorest performance, with $P_\text{D}$ dropping below 0.5 under high throughput and low false alarm conditions. These results clearly demonstrate that DCFNet achieves a superior trade-off between sensing reliability and communication efficiency, establishing a Pareto-optimal boundary in the field of FFT-based ISAC systems.

\subsection{Estimation Performance of DCFNet Compared to Conventional Detection methods}

In this subsection, we evaluate the range and velocity RMSEs (root mean squared errors) of DCFNet, conventional ML, ESPRIT, and FFT methods. We assume a harsh environment: two sensing targets move at the same Doppler bin, $\bar{f}_{\text{D}} = 0.95$ (corresponding to $v \approx 23.19~\text{m/s}$). The stronger target signal is received at $\text{SNR}=20~\text{dB}$, whereas the weaker target's SNR is swept from $-20~\text{dB}$ to $10~\text{dB}$. 
Because the two targets occupy the same Doppler bin, the Doppler-induced ICI generated by the strong reflector spreads across neighbouring subcarriers and raises the noise floor at the weak target's range cell, effectively burying its echo.

\begin{figure}[t]
    \centering
    \includegraphics[width=\linewidth]{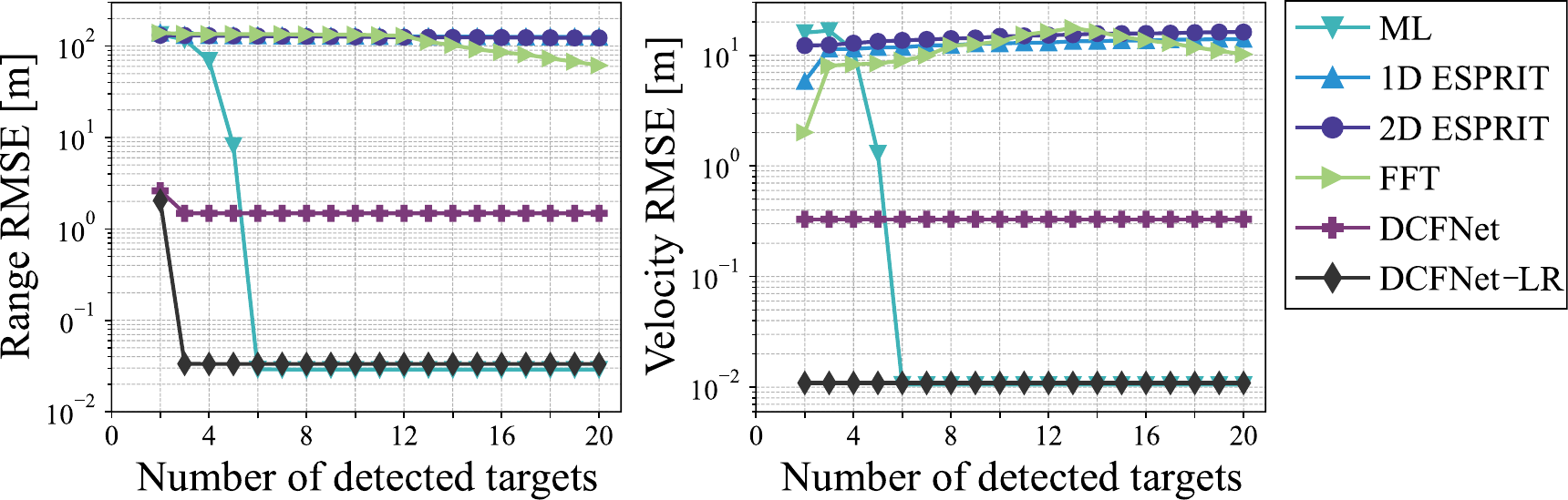}
    \vspace{-0.2cm}
    \caption{Range and velocity RMSE performance of DCFNet and the baseline methods as a function of the number of detected targets. The SNR of the weak target is set to $-10~\text{dB}$.}
    \label{fig:range_RMSE_per_target}
    \vspace{-8pt}
\end{figure}

In Fig.~\ref{fig:range_RMSE_per_target}, we fix the weak target SNR at $-10~\text{dB}$ and varies the effective number of detections processed by each algorithm. For FFT, the number of detected targets is controlled by the CA-CFAR threshold: a single target detection corresponds to a lower false-alarm probability, whereas twenty target detections reflect a higher false-alarm probability that admits many sidelobe and clutter responses. ML and ESPRIT do not employ CFAR; instead they are set to recover the strongest $C$ echoes, with $C$ swept from two to twenty. Since DCFNet produces a confidence map over the entire RV map, the cells are sorted by confidence, and only the top $C$ are forwarded. 
% When just one detection is processed, all methods first detect the strong target, and thus their range and velocity RMSEs of the weak target are low and nearly indistinguishable. As $C$ grows, the probability that the weak target is present rises.

Due to the ICI, the FFT detector fails to recover the weak target when twenty detections are processed. Similarly, ESPRIT, operating under a severely low SNR, is dominated by the strongest scatterer, and thus fails to detect the weak echo. The exhaustive ML search is able to identify the weak target after six iterations of the GLRT procedure; accurate range and velocity estimates require multiple rounds of strong-target cancellation. In contrast, DCFNet successfully suppresses the ICI from both targets and assigns high confidences to the weak echo even when it is deeply buried beneath the sidelobes of the strong one. As a result, its RMSE converges to the minimum value within the three target detections. Incorporating the local refinement (DCFNet-LR) tightens this curve further, achieving sub-meter accuracy by $C=3$. Therefore, DCFNet and DCFNet-LR reach the low-RMSE regime with significantly fewer detections while detecting the weak target, enabling a much lower false-alarm probability compared to the baseline approaches.

\begin{figure}[t]
    \centering
    \includegraphics[width=0.65\linewidth]{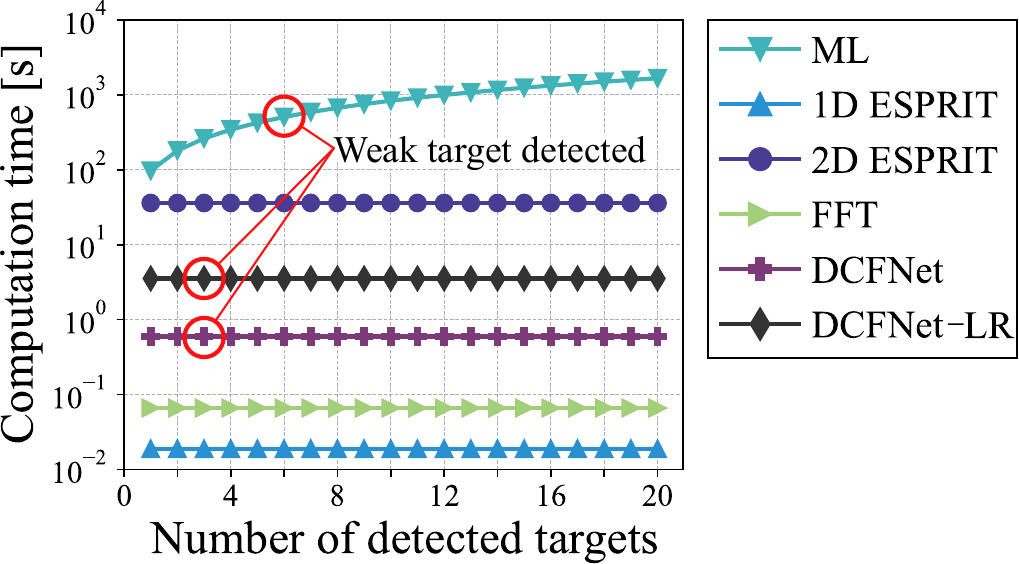}
    \vspace{-0.2cm}
    \caption{Execution time per various number of detected targets. The red mark represents when the weak target is detected. ESPIRIT and FFT fail to find the weak target.}
    \label{fig:range_RMSE_execution_time}
    \vspace{-0.2cm}
\end{figure}

In Fig.~\ref{fig:range_RMSE_execution_time}, although the ML-based method achieves fine RMSE performance for range and velocity, its compute cost remains prohibitively high. Moreover, while both ESPRIT and ML offer moderate runtime performance, they fail to reliably detect the weak target. In contrast, DCFNet is capable of detecting the weak target with significantly lower compute time than ML. Building upon this, DCFNet-LR refines the coarse estimates produced by DCFNet by restricting the search space, thereby reducing the computation time by approximately a factor of 175 while maintaining high estimation accuracy.

\begin{figure}[t]
    \centering
    \includegraphics[width=\linewidth]{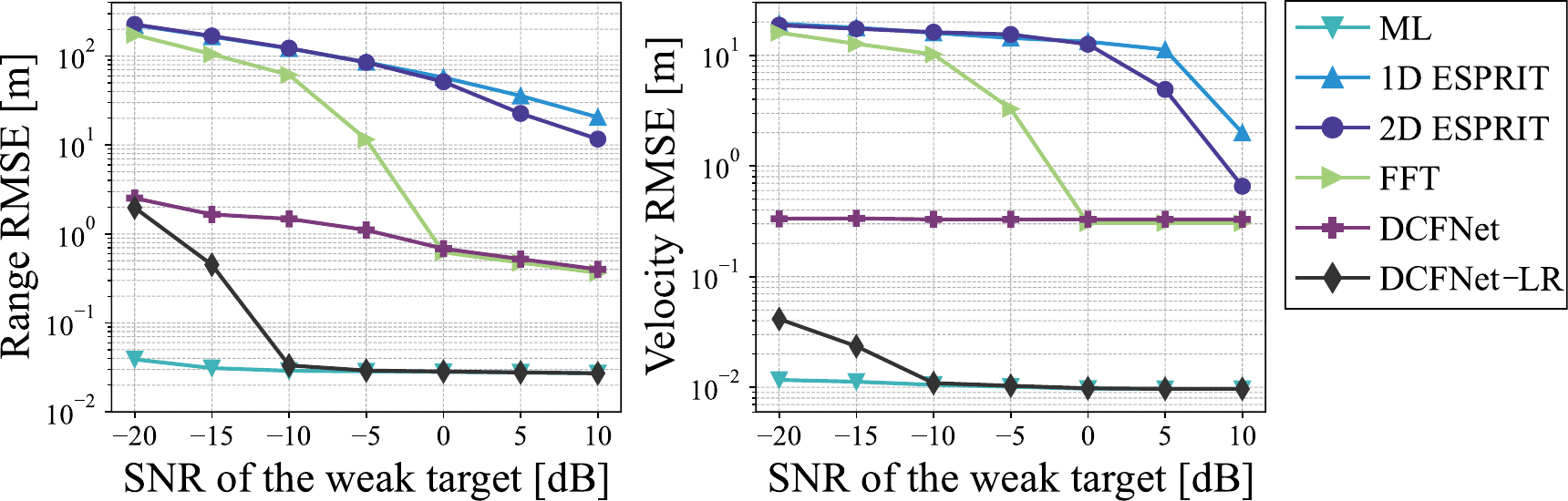}
    \vspace{-0.2cm}
    \caption{Range and velocity RMSE performance of the detection methods with respect to the SNR of the weak target.}
    \label{fig:range_vel_RMSE}
    \vspace{-8pt}
\end{figure}

In Fig.~\ref{fig:range_vel_RMSE}, the conventional FFT detector---constrained by sub-carrier spacing---fails once the weak echo drops below approximately $-5~\text{dB}$. 
Doppler-induced ICI raises the sidelobe floor, leading to a sharp increase in both range and velocity RMSE.
While ESPRIT provides super-resolution capabilities, it relies heavily on high SNR and a large number of snapshots; consequently, its performance degrades significantly in low-SNR regimes.
ML achieves near-optimal accuracy via exhaustive search but suffers from high computational complexity, which worsens with finer sub-grids, making real-time deployment impractical.
In contrast, DCFNet maintains range errors below $3~\text{m}$ and velocity errors below $0.3~\text{m/s}$ by learning to suppress ICI and compensate ICI offsets.
Incorporating a lightweight local GLRT refinement around each coarse DCFNet estimate (DCFNet-LR) confines the search to a $64 \times 64$ neighborhood, achieving sub-meter range and centimeter-per-second velocity accuracy, while reducing runtime by approximately two orders of magnitude compared to full-grid ML.

\section{Conclusion}
We propose DCFNet, a DL solution that simultaneously achieves high-resolution sensing, low computation cost, and effective ICI suppression in OFDM MU-MIMO ISAC systems. By shifting Doppler-induced ICI away from critical velocity regions and cancelling the residual interference, DCFNet produces clean RV maps without compromising communication throughput. A lightweight GLRT refinement module then upgrades the coarse detections to sub-cell accuracy with minimal latency. Simulations confirm that DCFNet shows robust performance in low-SNR, high-interference scenarios and reliably detects sensing targets with lower computation time compared to the conventional FFT, ESPRIT, and ML detection methods.

\bibliographystyle{IEEEtran}
\bibliography{{IEEEabrv,radcomm}}

\end{document}